\documentclass[bibyear]{aa}  
\usepackage[T1]{fontenc}

\makeatletter
\renewcommand*\aa@pageof{, page \thepage{} of \pageref*{LastPage}}
\makeatother

\usepackage{graphicx}
\usepackage{txfonts}
\usepackage{natbib}
\bibpunct{(}{)}{;}{a}{}{,} 
\usepackage{booktabs}
\usepackage{multicol}
\usepackage{graphicx}
\usepackage{fancyhdr}
\usepackage[hidelinks,colorlinks=true,linkcolor=blue,citecolor=blue,urlcolor=blue]{hyperref}
\usepackage{amsmath}	
\usepackage{amssymb}	
\usepackage[inter-unit-product=\cdot]{siunitx}
\usepackage{enumerate}
\usepackage{multirow}
\usepackage{mathtools}
\usepackage{comment}
\usepackage{longtable}
\usepackage{tabularx}
\usepackage{xcolor} 
\usepackage{ulem} 
\usepackage{comment}
\usepackage{cuted}

\newcommand{\Msun}{\,\mathrm{M}_\odot}
\newcommand{\Rsun}{\,\mathrm{R}_\odot}
\newcommand{\MSMBH}{M_\mathrm{SMBH}}
\newcommand{\tonde}[1]{\left(#1\right)}
\newcommand{\quadre}[1]{\left[#1\right]}
\newcommand{\graffe}[1]{\left\{#1\right\}}

\newcommand{\average}[1]{\left\langle#1\right\rangle}

\newcommand{\fastcluster}[1]{{\sc fastcluster}}

\DeclareSIUnit\year{yr}
\DeclareSIUnit\au{AU}
\DeclareSIUnit\parsec{pc}
\DeclareSIUnit\erg{erg}

\defcitealias{SG}{SG}

\newcommand{\orcidicon}[1]{\href{https://orcid.org/#1}{\includegraphics[width=11pt]{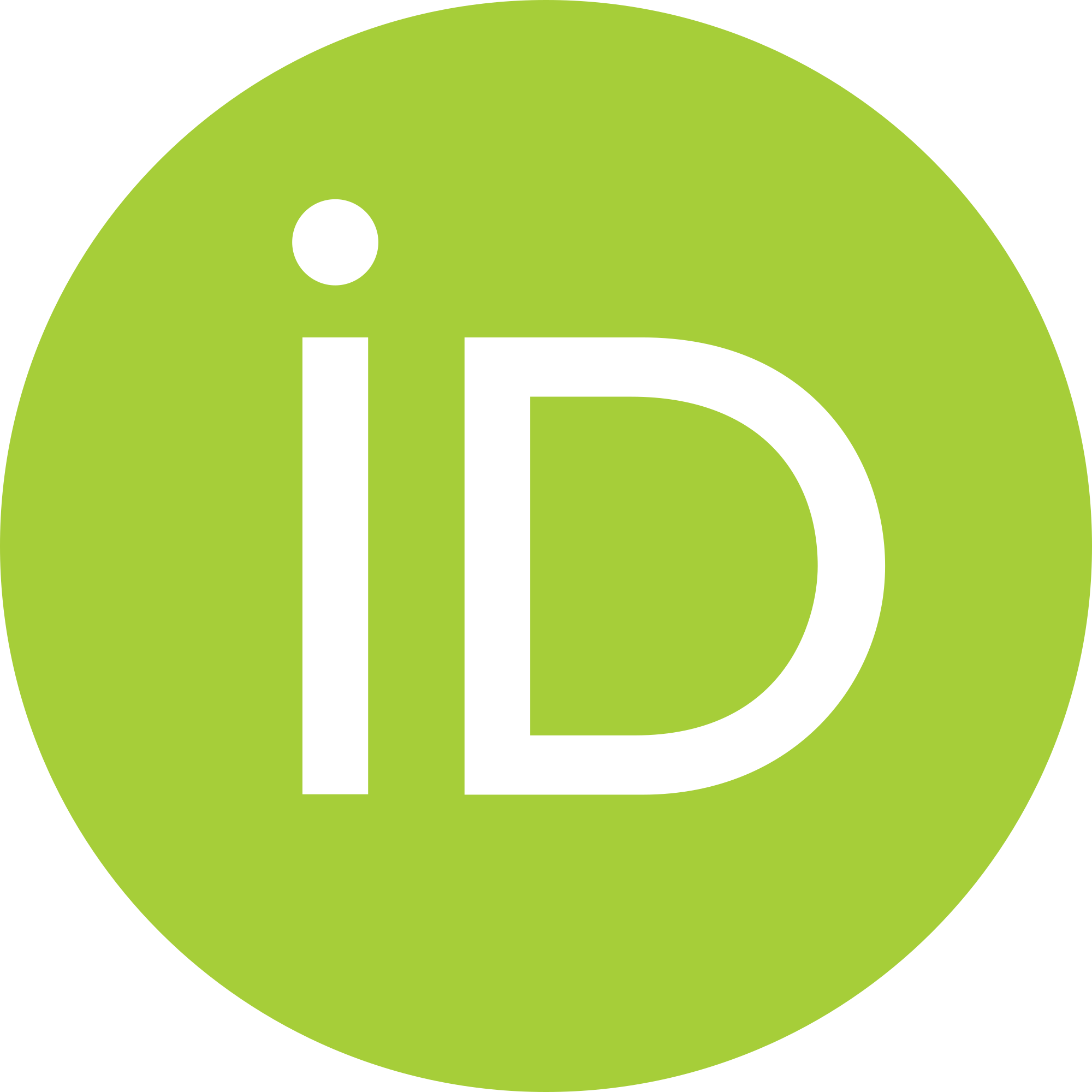}}}
\newcommand{\orcid}[1]{\href{https://orcid.org/#1}{\protect\orcidicon{#1}}}

\begin{document} 

\title{Impact of gas hardening on the population properties of hierarchical black hole mergers in AGN disks}
\titlerunning{Impact of gas hardening on hierarchical BH mergers in AGN disks}

\author{
M. Paola Vaccaro,\inst{1,2}\orcid{0000-0003-3776-9246}\thanks{\href{mailto:mariapaolavaccaro@gmail.com}{mariapaolavaccaro@gmail.com}}
Michela Mapelli,\inst{1,2,3,4}\orcid{0000-0001-8799-2548}\thanks{\href{mailto:mapelli@uni-heidelberg.de}{mapelli@uni-heidelberg.de}}
Carole Périgois,\inst{2,3}\orcid{}
Dario Barone,\inst{2}\orcid{0009-0008-9910-3961}\\
M. Celeste Artale,\inst{5}\orcid{0000-0003-0570-785X}
Marco Dall'Amico,\inst{2,3}\orcid{0000-0003-0757-8334}
Giuliano Iorio\inst{2,3,4}\orcid{0000-0003-0293-503X}
\and Stefano Torniamenti\inst{1,2,3,4}\orcid{0000-0002-9499-1022}}
\authorrunning{M. P. Vaccaro et al.}
\institute{
$^{1}$ Institut f{\"u}r Theoretische Astrophysik, ZAH, Universit{\"a}t Heidelberg, Albert-Ueberle-Stra{\ss}e 2, D-69120, Heidelberg, Germany\\
$^{2}$ Physics and Astronomy Department Galileo Galilei, University of Padova, Vicolo dell’Osservatorio 3, I–35122, Padova, Italy\\
$^{3}$ INFN–Padova, Via Marzolo 8, I–35131 Padova, Italy\\
$^{4}$ INAF–Osservatorio Astronomico di Padova, Vicolo dell’Osservatorio 5, I–35122, Padova, Italy\\
$^{5}$ Departamento de Ciencias Fisicas, Universidad Andres Bello, Fernandez Concha 700, Las Condes, Santiago, Chile}

\date{Received November 6, 2023}

 
\abstract{
    Hierarchical black hole (BH) mergers in active galactic nuclei (AGNs) are unique among formation channels of binary black holes (BBHs) because they are likely associated with electromagnetic counterparts and can efficiently lead to the mass growth of BHs. 
    Here, we explore the impact of gas accretion and migration traps on the evolution of BBHs in AGNs.
    We have developed a new fast semi-analytic model, which allows us to explore the parameter space while capturing the main physical processes involved. 
    We find that effective exchange of energy and angular momentum between the BBH and the surrounding gas (hereafter, gas hardening) during inspiral 
    greatly enhances the efficiency of hierarchical mergers, leading to the formation of intermediate-mass BHs (up to $10^4 \Msun$) and triggering spin alignment. Moreover, our models with efficient gas hardening 
    show both an anti-correlation between BBH mass ratio and effective spin, and a correlation between primary BH mass and effective spin.
    In contrast, if gas hardening is inefficient, the hierarchical merger chain is already truncated after the first two or three generations. 
    We compare the BBH population in AGNs with other dynamical channels as well as isolated binary evolution.}

\keywords{gravitational waves -- black hole physics -- stars: black holes -- stars: kinematics and dynamics -- galaxies: nuclei -- galaxies: active}

\maketitle
 
\section{Introduction}
\label{sec:introduction}
The first direct detection of gravitational waves (GWs) 
in 2015 
\citep{GW150914} 
has paved the ground for the study of binary black holes (BBHs). 
More than 90 GW event candidates 
have been detected to date, most of them associated with BBHs \citep{GWTC3_first,GWTC3_second}. 
A few 
BBH candidates like GW190521 \citep{GW190521} and possibly GW190403\_051519 and GW190426\_190642 \citep{GWTC2.1, GWTC3_first, GWTC3_second} stand out among the other detections because they involve 
black holes (BHs) in the pair-instability mass gap, 
challenging traditional models of stellar evolution \citep{Woosley_2002, Woosley_2014, Woosley_2021, Belczynski_2016, sevn_2017, Stevenson_2019, OBrien_2021, Siegel_2022, Sabhahit_2023, Umeda_2023} and raising questions about their formation \citep{Farmer_2019, Farmer_2020, sevn_2020, Belczynski_2020, Marchant_2020, Costa_et_al_2020, Farrell_2021, Vink_2021, Tanikawa_2021, Tanikawa_2022, DallAmico_2021, Banerjee_2022, Moreno_Mendez_2023}. 

Stellar dynamics provides some of the most straightforward channels to explain the birth of such oversized BHs, via star-star collisions \citep{DiCarlo_2019, DiCarlo_2020, Kremer_2020, Renzo_2020, torniamenti2022, Costa_2022, Ballone_2023}, 
or repeated mergers of stellar-origin black holes 
(\citealt{Miller_2002,Fishbach_2017,Gerosa_2017,Rodriguez_2019,Doctor_2020,Kimball_2020,Flitter_2021}; see, e.g., \citealt{Mapelli2021_review} for a review). The latter process, often called hierarchical mergers, 
takes place only in dense star clusters, where merger remnants can be retained inside the system 
\citep[e.g.,][]{Antonini_2019,Fragione_t3bb} and pair up again with other single BHs via dynamical  
encounters \citep[e.g.,][]{Heggie_hard_bin, Zwart_2000}. In order to constrain the origin of the observed BBH mergers, it is important to characterize the hierarchical merger process in different environments such as young star clusters \citep[YSCs, e.g.,][]{Ziosi_2014, Mapelli_2016, Banerjee_2017, Banerjee_2017b, Banerjee_2020, DiCarlo_2020, Kumamoto_2019, Kumamoto_2020}, globular clusters \citep[GCs, e.g.,][]{Downing_2010, Rodriguez_2015, Rodriguez_2016, Rodriguez_2018, MOCCA_2017, Fragione_Kocsis_2018, Zevin_2019, Antonini_Gieles_2020, Antonini_Gieles_2023}, nuclear star clusters \citep[NSCs, e.g.,][]{O'Leary_2009, Miller_2009, Antonini_2016, Petrovich_2017, Leigh_McKernan_2018, ArcaSedda_2018, ArcaSedda_2020b, Atallah_2023, Chattopadhyay_2023}, and active galactic nuclei  \citep[AGNs, e.g.,][]{McKernan_2012, McKernan_2020, McKernan_2022, Bartos_2017, Stone_2017, Yang_2019, Secunda_2020, Tagawa_2020, Tagawa_2020_b, Tagawa_2022, Tagawa_2023, Ford_2022}. 

In AGNs, a central supermassive black hole (SMBH) is surrounded by a dense gaseous accretion disk. 
AGNs can appear as extremely luminous objects called quasars, but also as Seyfert galaxies, radio galaxies, or blazars, depending on their luminosity and our viewing angle (see \citealt{Netzer_AGN_review} for a review of the unification scheme and its controversy). Stars and stellar-sized BHs orbiting the SMBH are subject to  
gas torques, that can bend their orbits aligning them to the disk. This is expected to lead to a large overdensity of BHs with similar orbits in small areas of the disk called migration traps \citep{McKernan_2012, Bellovary_2016}, 
where BHs can pair-up efficiently via gas capture \citep{DeLaurentis_2023,Li_2023,Rowan_2023,Rowan2_2023,whitehead_2023}. Therefore, AGNs are 
potential factories of BBH mergers with mass in the pair-instability mass gap and above \citep{Yang_2019, Tagawa_2020_b, Tagawa_2020}. 

This channel has attracted great interest because GW signals from mergers could be accompanied by some electromagnetic emission \citep{Bartos_2017, Tagawa_2023}, possibly anticipated by a neutrino detection \citep{zhu2023_neutrinos,zhou2023_neutrinos}. To explore this possibility, electromagnetic follow-up observations have been carried out for
 several GW events
but so far all associations with BBH mergers remain controversial \citep{Greiner_2016, Coughlin_2020, bustillo2021_gw190521}.

Several studies investigate the possibility that the high-mass BBH merger event GW190521 is associated with an AGN disk \citep{Tagawa_2021, Samsing_2022} because of its large mass, high spin \citep{GW190521} and claimed 
electromagnetic counterpart \citep{Graham_gw190521, bustillo2021_gw190521, Morton_in_prep}. \cite{Yang_2019} explore the AGN scenario for another high-mass BBH, GW170729, which has support for non-zero effective and precessing spins. Some authors such as \citet{Gayathri_2021, Ford_2022} have predicted that a sizeable fraction ($\sim 20\%$ up to $80\%$) of the observed BBH mergers may originate in AGNs, while an analysis based on the sky localization of GW signals reveals that the fraction of detected BBH mergers originated in bright ($L\geq10^{46}\,\si{\erg}\, \si{\second^{-1}}$) AGNs cannot be higher than $17\%$ \citep{Veronesi_2023}.

BBH pair-up in AGNs can happen either in the migration traps or at other locations in the disk \citep{Wang_2021}, commonly referred to as `the bulk'. \citet{McKernan_2020} find that, although more than $50\%$ of mergers happen in the bulk, hierarchical mergers are only efficient in migration traps. Moreover, \citet{Tagawa_2020} find that BBHs assembled in the bulk via gas capture migrate toward the migration trap while hardening.

The complex physics of BBHs in AGN disks has been extensively explored in previous literature, accounting for the effects of binary-single interactions \citep{Stone_2017, Leigh_McKernan_2018} and gas torques (using models borrowed from proto-planetary physics, e.g. \citealt{McKernan_2012, Bartos_2017,Yang_2019,Yang_2019_bis,Secunda_2020}; or hydrodynamical simulations, e.g. \citealt{Li_2023, Kaaz_2023, Rowan_2023,Rowan2_2023, whitehead_2023}). For example, \citet{Ishibashi_2020} explore the effects of energy and angular momentum exchange between the BBH and the surrounding gas. 
In particular, they assume that the BBH evolves surrounded by a circumbinary disk, which induces orbital decay and pumps its orbital eccentricity (hereafter, gas hardening).

Here, we study hierarchical BBH mergers in the migration traps of AGN disks and compare our results to 
mergers in  YSCs, GCs, and NSCs \citep{fastcluster2021,fastcluster2022}.  
Specifically, we test the impact of gas hardening \citep{Ishibashi_2020} on the BBH merger population in AGNs. We find that the hierarchical merger process is pronouncedly more efficient when 
accounting for gas hardening. 

We have developed a new semi-analytic code for the simulation of hierarchical mergers in AGNs, which is effective in exploring the parameter space, and models the relevant physical processes while being much faster than an N-body or hydrodynamical code. The new code is publicly available inside the \fastcluster{} software environment \citep{fastcluster2021,fastcluster2022}.



\begin{figure*}
    \centering
    \includegraphics[width=\linewidth]{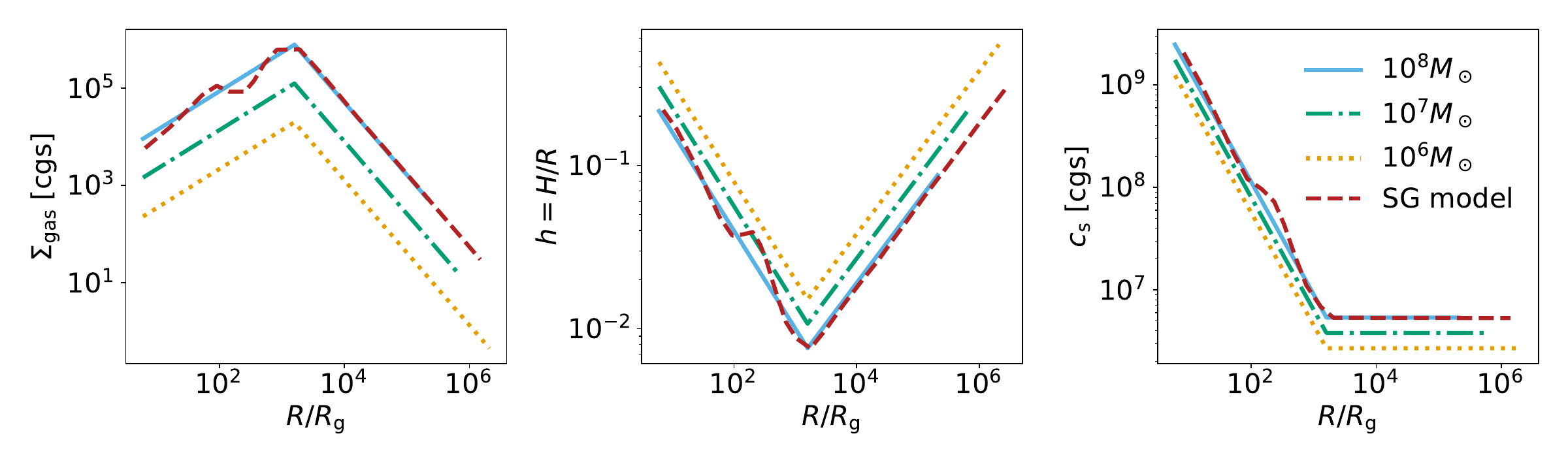}
    \caption{The figure shows the surface density $\Sigma_\mathrm{gas}$, aspect ratio $h$ and sound speed $c_\mathrm{s}$ profiles as a function of the scale distance $R/R_\mathrm{g}$. Maroon dashed lines represent the \citetalias{SG} model for a disk with viscosity $\alpha=0.01$ around a $10^8\Msun$ SMBH. Blue solid lines display our broken power-law best fit to the \citetalias{SG} model for $\alpha=0.01$ and $\MSMBH=10^8\Msun$. The green dash-dotted and orange dotted lines show our fits for a $10^7\Msun$ and a $10^6\Msun$ SMBH, respectively (both with $\alpha=0.01$).}
    \label{fig:SG-model}
\end{figure*}

\section{Methods}


\subsection{AGN disk model}
\label{sec:setup_AGN_methods}
We assume the AGN disk to be described by  \cite{SG} (hereafter, \citetalias{SG}). These authors introduce a hydro-dynamical model for a geometrically thin and optically thick disk with steady-state accretion of $0.1\, \dot{M}_\mathrm{Edd}$ onto the central SMBH. Gas turbulence is 
assumed to be the cause of disk viscosity, characterized by the viscosity coefficient $\alpha \in \quadre{0,\,1}$. This model neglects any effects due to magnetic fields and general relativity.
The physical parameters of the disk are functions of the mass of the SMBH $M_{\rm SMBH}$, the distance $R$ from the SMBH, and the viscosity parameter $\alpha$. We slightly simplify the radial dependence of the \citetalias{SG} model for numerical ease and re-scale their functions to allow for different $\MSMBH$ and $\alpha$ parameters as shown in Fig.~\ref{fig:SG-model}.
The resulting expressions for the gas surface density $\Sigma_\mathrm{gas}$, the disk aspect ratio $h=H/R$ (defined as the ratio between the height of the disk $H\tonde{R}$ and its radius $R$), and the sound speed $c_\mathrm{s}$ are the following.

\begin{strip}
    \rule{\dimexpr(0.5\textwidth-0.5\columnsep-0.4pt)}{0.4pt}%
\begin{equation}
    \Sigma_\mathrm{gas}\tonde{R,\MSMBH,\alpha} = \SI{7.94e5}{\gram \, \centi \meter^{-2}} \tonde{\frac{\MSMBH}{10^8 \Msun}}^{4/5} \tonde{\frac{\alpha}{0.01}}
    \begin{dcases}
    \tonde{\frac{R}{10^{3.2}\,R_\mathrm{g}}}^{0.8} \quad &R \leq 10^{3.2}\,R_\mathrm{g}\\
    \tonde{\frac{R}{10^{3.2}\,R_\mathrm{g}}}^{-1.49} \quad &R > 10^{3.2}\,R_\mathrm{g}
    \end{dcases}
    \label{eq:Bartos2017, Sigma gas} 
\end{equation}
\begin{equation}
    h\tonde{R,\MSMBH,\alpha} = \SI{7.59e-3}{} \tonde{\frac{\MSMBH}{10^8 \Msun}}^{-3/20} \tonde{\frac{\alpha}{0.01}}^{-1}
    \begin{dcases}
    \tonde{\frac{R}{10^{3.2}\,R_\mathrm{g}}}^{-0.6} \quad &R \leq 10^{3.2}\,R_\mathrm{g}\\
    \tonde{\frac{R}{10^{3.2}\,R_\mathrm{g}}}^{0.5} \quad &R > 10^{3.2}\,R_\mathrm{g}
    \end{dcases}
    \label{eq:Bartos2017, H} 
\end{equation}
\begin{equation}
    c_\mathrm{s} \tonde{R,\MSMBH,\alpha}= \SI{5.37e6}{\kilo \meter \, \second^{-1}} \tonde{\frac{\MSMBH}{10^8 \Msun}}^{3/2} \tonde{\frac{\alpha}{0.01}}^{-1}
    \begin{dcases}
    \tonde{\frac{R}{10^{3.2}\,R_\mathrm{g}}}^{-1.1} \quad &R \leq 10^{3.2}\,R_\mathrm{g}\\
    \quad 1 \quad &R > 10^{3.2}\,R_\mathrm{g}
    \end{dcases}
    \label{eq:Bartos2017, c sound}
\end{equation}
\hfill
    \rule{\dimexpr(0.5\textwidth-0.5\columnsep-0.4pt)}{0.4pt}
\end{strip}
In the above equations, we define the gravitational radius as $R_\mathrm{g}=G M_{\rm SMBH}/c^2$, where $G$ is the gravity constant and $c$ the speed of light.

The viscosity coefficient $\alpha$ is a free parameter. We assume it to be constant over the whole extension of the disk and to have a constant value independent of the other physical properties of the disk. The assumed value is $\alpha =0.1\,$ consistent with observations \citep{alpha_viscosity}.

Migration traps are locations in the disk where 
migration stalls and BHs pile up. \cite{Bellovary_2016} find that migration traps are found at locations where the slope of the gas surface density profile changes sign from positive to negative, i.e. at local maxima. 
In a \citetalias{SG} disk there are two local maxima in the gas surface density and therefore two migration traps: an inner trap at ${\approx 100}\, R_\mathrm{g}$ and an outer one at ${\approx 1300}\, R_\mathrm{g}$. 


We notice from Fig.~\ref{fig:SG-model} 
that the inner migration trap coincides with a local maximum in the density profile of the original \citetalias{SG} model, while the outer migration trap corresponds to a global maximum. For simplicity, we ignore the local overdensity in the disk at $10^2\, R_\mathrm{g}$ and we only assume the existence of the outer migration trap. We thus define the trap radius as $R_\mathrm{trap} =$ $10^{3.1}\, R_\mathrm{g}$, although its location and existence may be affected by other effects \citep[e.g.][more details in Section~\ref{sec:discussion_migration}]{grishin2023, Pan_2021}.

The disk is assumed to have radial extension between the innermost stable circular orbit (ISCO) radius for a non-rotating BH, which we call $R_\mathrm{min}=6\,{}R_\mathrm{g}$ in this context, and an outer radius $R_\mathrm{max}=\SI{0.1}{\parsec}\,{}(M_{\rm SMBH}/10^6\Msun)^{1/2}$, beyond which the disk's self-gravity becomes important \citep{Goodman_2003, Yang_2019_bis}. For $R>R_\mathrm{max}$, the disk is expected to fragment and experience star formation. Energetic feedback from the newly-formed stars may keep the disk vertically supported \citep{SG}, but the 
outcome of viscous interactions with the BHs in this region of the disk is highly uncertain, so we conservatively neglect it. Moreover, we neglect the contribution of stars formed in AGN disks, 
which are expected to be massive 
and whose compact remnants might also become GW sources \citep{Cantiello_2021, wang2023_starformationAGN}.

We assume a non-spinning SMBH. 
We randomly generate the SMBH mass from the observational distribution derived by \citet{Greene_2007} in the Local Universe, that is a 
Gaussian distribution with mean $\langle{}\log{M_{\rm SMBH}/\Msun{}}\rangle{}=6.576\pm{} 0.591$. 


A mass-accretion episode onto a SMBH lasts for a finite amount of time, which we refer to as the AGN disk lifetime. The lifetime of AGN disks is subject to large uncertainty:  
different estimates span several orders of magnitude in the range of $10^{-2} - 10^3\, \si{\mega\year}$ \citep{Khrykin_quasar_lifetime}.
Also, it is not clear whether accretion onto SMBHs happens continuously over a given time span, or episodically through many cycles of efficient accretion. Here, we use the estimate by \citet{Khrykin_quasar_lifetime}, based on observations of quasars' proximity effect. They find that the quasar lifetime $\tau$ is distributed according to a Gaussian distribution with mean $\langle{}\log{\tau/{\rm Myr}}\,\rangle{}=0.22\pm{}0.80$. 
We randomly extract the lifetime for AGN disks in our model from this distribution.

There are other models for stable AGN accretion disks, with notably different features compared to the one we adopt here. 
For example, the model by \cite{TQM} 
differs from the SG model in both density and aspect ratio 
\citep{McKernan_2022}. We will explore the impact of different disk models in a follow-up study.

\subsection{Nuclear star cluster (NSC)}
\label{sec:NSC parameters}
SMBHs and NSCs  commonly inhabit galactic spheroids  
with stellar masses ranging from $10^8\Msun$ to $10^{11}\Msun$ \citep{Graham_Spitler}. For  $\MSMBH \leq{} \SI{5e7}{\Msun}$, the mass of the SMBH and that of the NSC scale as 
\begin{equation}
    \log{\tonde{\frac{\MSMBH}{\MSMBH+M_\mathrm{NSC}}}} = 
    \frac{2}{3}\,{} \log{\tonde{\frac{\MSMBH}{\SI{5e7}{\Msun}}}} 
    \label{eq: SMBH-NSC-Graham-Spitler}
\end{equation}
For $\MSMBH > \SI{5e7}{\Msun}$, the SMBH generally does not coexist with a NSC and we set $M_\mathrm{NSC}=0$.

The effective radius of a NSC mildly correlates to its mass \citep{Neumayer_2020}. The best-fit relation between the mass of the NSC and its effective radius is
\begin{equation}
    \log\tonde{\frac{r_\mathrm{h}}{\si{\parsec}}} = 
    \begin{dcases}
    0.538 & M_\mathrm{NSC} < 10^6 \Msun \\
    0.228\, \log\tonde{\frac{M_\mathrm{NSC}}{\Msun}} -1.19 & M_\mathrm{NSC} \geq 10^6 \Msun
    \end{dcases}
    \label{eq:r_h_NSC}
\end{equation}

To account for the spread in the data, we sample $\log\tonde{r_\mathrm{h}/\si{\parsec}}$ from a Gaussian distribution with mean centered on the corresponding value determined by eq.~\ref{eq:r_h_NSC} and width $\sigma = 0.1$ ($0.2$) for $M_\mathrm{NSC} <10^6\Msun$ ($\geq10^6\Msun$), so that most of the data fall under the $\pm 2\sigma $ dispersion.

We approximate the spatial distribution of stars in the NSC with a \cite{plummer1911} model, with mass 
\begin{equation}
    M_\mathrm{NSC}(R) = M_\mathrm{NSC}\, \frac{R^3}{\tonde{R^2+a_\mathrm{PL}^2\,}^{3/2}}
    \label{eq:Plummer integrated}
\end{equation}
where $R$ is the distance from the SMBH, and $a_\mathrm{PL} = r_\mathrm{h}/(\SI{1.3}{\parsec})$ is the scale parameter for the Plummer model.

The mass fraction of stellar-origin BHs in the AGN disk $f_\mathrm{BH}$ is sampled from a Gaussian with mean 0.04 and standard deviation 0.01 to account for mass segregation in the NSC \citep{Bartos_2017}. 
 The number of BHs in the AGN disk and their cumulative mass are thus given by the following equations:
\begin{equation}
    N_\mathrm{BH}=\frac{f_\mathrm{BH}\, f_\mathrm{trap}}{2}\, \frac{M_{\rm NSC}(R_\mathrm{max})}{\average{m_\ast}}\,, \quad M_\mathrm{BH}^\mathrm{max} = N_\mathrm{BH} \average{m_\mathrm{BH}},
    \label{eq:N_BH_max}
\end{equation}
where $M_{\rm NSC}(R_{\rm max})$ is defined in eq.~\ref{eq:Plummer integrated}, the mean stellar-origin BH mass $\average{m_\mathrm{BH}}$ is computed from data obtained from the population synthesis simulation code \textsc{sevn} \citep{sevn_2023} at solar metallicity (Appendix~\ref{sec:oldfastcluster}), the distance $R_\mathrm{max}$ is the outer radius of the disk 
(Section~\ref{sec:setup_AGN_methods}), the factor $1/2$ accounts for prograde orbiters\footnote{Prograde orbiters are objects that orbit in the same direction as the disk. Here, we assume that half of the total BH population in the NSC are prograde orbiters.} only, 
and the mean stellar mass $\average{m_\ast}\simeq 1 \Msun$ is computed using a \cite{Kroupa_2001} initial mass function.

The parameter $f_\mathrm{trap}$ is the ratio 
between the number of BHs that are able to reach the migration trap on a timescale shorter than the disk lifetime $\tau$ and the total number of BHs that interact with the disk: this study focuses on BBH pair-ups in the migration trap, therefore we are not interested in any BHs that live outside of that location. For an operational definition of $f_\mathrm{trap}$, see Section~\ref{sec:pair-up}.

We assume the velocity dispersion of stars to scale with the SMBH mass as \citep{Merritt_2001}
\begin{equation}
    \sigma = \SI{200}{\kilo \meter\,{} \second^{-1}} \tonde{\frac{\MSMBH}{10^8\Msun}}^{1/5} .
    \label{eq: sigma velocity dispersion}
\end{equation}
However, it has been shown \citep[e.g.][]{Scott_Graham_2013, Sahu_2019, Graham_2023} that this result
depends on the morphology of galaxies included in the sample. For example, \citet{Sahu_2019} find an exponent $\sim 1/6$ and show that 
it is caused by the combined contribution of S\'ersic (i.e. following the \citealt{Sersic_1963} brightness profile) and core-S\'ersic (i.e. centrally depleted) galaxies following two
different $\MSMBH - \sigma$ relations with exponents $\tonde{5.75\pm0.34}^{-1}$ and $\tonde{8.64\pm1.10}^{-1}$ respectively.

\subsection{First-generation (\texorpdfstring{$1g$}{1g}) BHs}
\label{sec:setupofinitialbhpop}

We randomly draw first-generation ($1g$, i.e. stellar-origin) BH masses $m_{\rm BH}$ from a catalog obtained with the population synthesis code \textsc{sevn} \citep{sevn_2017, sevn_2019, sevn_2020, sevn_2023}. {\sc sevn} relies on up-to-date stellar tracks \citep{Bressan_2012, Costa_2019, Nguyen_2022} and models the formation of compact objects by taking into account electron-capture \citep{Giacobbo_2019}, core-collapse \citep{sevn_2023} and pair-instability supernovae \citep{sevn_2020}. In particular, here we assume the rapid core-collapse supernova model by \cite{Fryer_2012}, which enforces the existence of a mass gap between the maximum neutron star mass and the minimum BH mass \citep{Oezel_2010}. 
We use the fiducial model from \citet{sevn_2023} and consider single stellar evolution only, as described in Appendix~\ref{sec:oldfastcluster}. We assume metallicity $Z=0.02$, i.e. approximately solar, matching the typical metallicity at the center of massive galaxies in the Local Universe \citep{Gallazzi_2008}.

We randomly draw an initial radial position for each BH. The radial extension of the accretion disk is small compared to the typical dimension of a NSC, so we neglect mass segregation on this scale and consider the numerical density of objects at radii $R<R_\mathrm{max}$ to be uniform in radius:
$ R \sim \mathcal{U}\tonde{R_\mathrm{min}, R_\mathrm{max}} $.

We draw the dimensionless spin magnitude $\chi$ from a Maxwellian distribution with one-dimensional root-mean-square $\sigma_\chi =0.1$, truncated at $\chi=1$. We choose $\sigma_\chi = 0.1$ because it is reminiscent of the spins inferred from the third GW transient catalog (GWTC-3, \citealt{GWTC3_second}). This assumption does not take into account that the BBH population in GWTC-3  likely comes from multiple formation channels, including the AGN disk scenario. Current data are not sufficiently informative to differentiate between formation channels. 
We set the primary spin tilt as in Appendix \ref{sec:spin_tilt}.

\subsubsection{Gas Capture}
\label{sec:damping}
After setting up the properties of $1g$ BHs, we follow their evolution in the disk. 
When NSC objects orbit around the central SMBH, their orbits can cross the disk and gather some of the disk gas, causing them to be subject to strong gas drag. This is expected to dampen both the inclination $i$ and the eccentricity $e$ of their orbit \citep{Cresswell_2007}. Therefore, after a sufficient number of laps, these objects will have circular orbits embedded in the disk. This process is called gas capture or orbital damping.

The gas accretion and subsequent gas drag are significant only for prograde orbiters, hence we neglect any variation in the orbits of retrograde orbiters.

We define the inclination damping timescale $t_\mathrm{damp}$ for a BH of mass $m_{\rm BH}$ on an initial orbit of semi-major axis\footnote{We often use coordinates such as semi-major axes and radii. We use capital letters $A$ and $R$ to refer to orbits around the central SMBH while we use lower-case letters $a$ and $r$ for orbits inside a binary system.} $A$ around a SMBH of mass $\MSMBH$ as \citep{wang2023}
\begin{equation}
    t_\mathrm{damp }\simeq \frac{c^4}{18\,{} G\,{} \Sigma_\mathrm{gas} \,{}m_{\rm BH}}\,{}\sqrt{\frac{A^3}{G \tonde{\MSMBH +m_{\rm BH}}}} ,
    \label{eq:t_damp}
\end{equation}
where 
$\Sigma_\mathrm{gas}\tonde{A}$ is the surface density of the gas. In this model, we neglect the mass increase due to gas accretion. Hence, during the orbital evolution, $m_\mathrm{BH}$ is a constant quantity.

\subsubsection{Migration}
\label{sec:migration_timescale}
Once a BH is embedded in the disk, it exchanges angular momentum with the surrounding gas and is subject to gas torques. Torques can be both positive or negative, leading to outward or inward migration, respectively. Similar to what happens to planet seeds in protoplanetary disks, migration can happen in two different ways called Type I and Type II. 

Small to medium-mass objects are subject to \emph{Type I migration}, meaning that they change their radial position in the disk without significantly perturbing the density distribution of the disk itself. 
For a BH of mass $m_{\rm BH}$ 
on a circular orbit with radius $R$, this happens on a timescale \citep{Lyra_2010,McKernan_2012,Paardekooper_migration}
\begin{equation}
    t_\mathrm{migr,\,I}= \frac{\MSMBH^2\,  h^2}{m_{\rm BH}\,{} \Sigma_\mathrm{gas} \,{}R^2\,{} \Omega} ,
    \label{eq:t_migr}
\end{equation}
where $h$ is the aspect ratio of the disk and $\Omega$ is the Keplerian angular velocity around the SMBH. 

Differently from eq. \ref{eq:t_damp}, here we are considering a radius $R$ rather than a semi-major axis $A$ because gas capture happens necessarily before migration,\footnote{Migration can only set-in when $i=0$ and the orbit is embedded in the disk. Because of gas drag, $i\to 0$ and $e\to 0$ on roughly the same timescale.} so the orbits have already been circularized when migration sets in.


In our disk model (Fig.~\ref{fig:SG-model}), torques are positive in the inner region of the disk, where the slope of the surface density is positive, and they are negative in the outer region, where the slope of the surface density is negative \citep{Bellovary_2016}. 
Therefore, Type I migration is directed outward in the inner disk and inward in the outer disk. At the location where the torques change sign, called a migration trap, migration will stall leading to a large accumulation of objects.
Hence, after a timescale $t_\mathrm{migr,\,I}$ (eq. \ref{eq:t_migr}), the migrating object will be in the migration trap. 

{\vskip 0.5cm}

Larger objects, on the other hand, can open gaps in the disk. This happens because the motion of a massive object exerts an intense tidal perturbation on the disk, which effectively pushes material away from the orbit's trail \citep{Bryden_1999}. This is called \emph{Type II migration}. An object of mass $m_\mathrm{BH}$ can open a gap in the disk if \citep{McKernan_2012_bis}
\begin{equation}
    q > \sqrt{\frac{\alpha}{0.09}\, h^5}, 
    \label{eq:open gap}
\end{equation}
where $q=m_{\rm BH}/\MSMBH$ is the mass ratio with respect to the central SMBH, $\alpha$ is the viscosity parameter, and $h=h\tonde{R}$ is the aspect ratio of the disk at radius $R$.


If an object opens a gap in the disk, assuming that no gas can cross the gap, its migration follows the viscous evolution of the disk's gas; 
hence the timescale for Type II migration is the timescale for the viscous evolution of the disk \citep{McKernan_2012}
\begin{equation}
        t_\mathrm{migr,\, II} = t_\mathrm{visc} = (\alpha \,{}h^2\,{} \Omega)^{-1}.
    \end{equation}

However, pressure forces in the disk push to close the gap. So, even if an object is massive enough to open a gap, the latter can stay open against pressure forces only if \citep{Bryden_1999, McKernan_2012_bis}
\begin{equation}
    q \gtrsim \alpha \tonde{40\,h}^2.  
    \label{eq:keep gap open}
\end{equation}

Type II migrators typically have high mass: taking as fiducial values $\alpha=0.1$ and $h=0.02$, eqs.~\ref{eq:open gap} and \ref{eq:keep gap open} 
entail $m_{\rm BH} \gtrsim 0.1\, \MSMBH$. They are bound to their radial location in the disk and can only move \emph{with} the disk on its viscous timescale, hence they will never reach the migration trap. Moreover, the gaps they create will prevent some Type I migrators from reaching the trap: they will intercept inward-moving migrators if they are located at a radius greater than the trap's, or they will intercept outward-moving migrators if the opposite is true. These intercepted BHs can potentially pair-up and merge with the Type II migrator, although their merger would not be assisted by gas hardening. However,
we expect such a massive BH in an AGN disk only in two cases: a  high-generation hierarchical BH or the central BH of a dwarf galaxy dragged into the AGN disk after a galaxy-galaxy merger \citep[e.g.,][]{dimatteo2008}. Describing galaxy mergers is out of the scope of this paper, while a high-generation hierarchical BH 
can form only in the late stages of the disk's lifetime. At that point, we can assume that most of the BHs have already migrated in the migration trap and we can neglect any further pair-up event with a Type II migrator.

Therefore, in our model we consider the onset of Type II migration to be one of the processes that can halt hierarchical mergers. Moreover, the presence of gaps has noteworthy consequences on disk structure. Nevertheless, we bluntly neglect any evolution of the disk density profile.

\subsubsection{Pair-up}
\label{sec:pair-up}

Depending on the physical properties of the AGN (such as viscosity, gas density, aspect ratio, and SMBH mass), gas capture and migration can happen on short timescales. When these processes are efficient, they can lead to a large abundance of BHs in the narrow region of the migration trap. Also, all BHs in the migration traps are on similar orbits (prograde and quasi-Keplerian), so their relative velocities of encounter are small. Under such conditions, it is easy for two BHs to become gravitationally bound in a binary. Therefore, efficient damping and migration lead to efficient binary pair-up.

In this work, we assume that the pair-up of a primary and secondary BH is immediate as soon as the primary reaches the migration trap. This assumption is fully justified by the efficiency of dynamical friction in the migration trap (see \citealt{Qian_2023} and Section~\ref{sec:discussion}). 
The pairing timescale of a BBH is therefore 
\begin{equation}
t_\mathrm{pair}=t_\mathrm{damp} + t_\mathrm{migr,\,I} + t_\mathrm{in},
\end{equation}
where $t_\mathrm{in}$ is the formation time of the primary BH since the time of formation of the disk. 

We consider each BBH to be in a circular orbit in the migration trap at a radius $R_\mathrm{trap}$ (Section~\ref{sec:setup_AGN_methods}). 
We compute the fraction $f_\mathrm{trap}$ of BHs that reach the migration trap by counting the number of $1g$ BHs for which $t_\mathrm{pair}<\tau$, where $\tau$ is the disk lifetime, and dividing it by the total number $N$ of simulated first-generation BHs. We use this parameter for the computation of the maximum mass that can be accreted by a single BH, as in eq.~\ref{eq:N_BH_max}.

\subsubsection{BBH properties}
We determine the secondary BH mass as in Appendix \ref{sec:secondary mass},
while we set the secondary BH spin in the same way as for the primary BH. We set the secondary BH spin tilt as in Appendix~\ref{sec:spin_tilt}. We assign the initial semi-major axis $a$ of the binary sampling from a distribution $p\tonde{a}\propto a^{9/2}$ \citep{Binney-Tremaine-2008, Tagawa_2020} for $a\in \quadre{a_\mathrm{min}, a_\mathrm{max}}$, where $a_\mathrm{min} = 1\Rsun$ and $a_\mathrm{max}$ is equal to the Hill radius, computed as
\begin{equation}
    R_\mathrm{Hill} = R_\mathrm{trap} \tonde{\frac{m_1+m_2}{3\, \MSMBH}}^{1/3},
    \label{eq:Hill_radius}
\end{equation}
where $m_1$ and $m_2$ are the primary and secondary BH mass, respectively. 
We set the initial eccentricity $e$ following a thermal distribution $p(e)\propto 2 e$ for $e$ between 0 and 1 \citep{Jeans_1919_binaries}.

According to the \cite{Heggie_hard_bin} law, a binary can survive in a star cluster only if it is hard, meaning that its binding energy 
is larger than the average kinetic energy of a field star
\begin{equation}
    E_b = \frac{G \,{}m_1\,{} m_2}{2\,{}a} \geq \average{E_k}= \frac{1}{2} \average{m_\ast}\sigma^2 ,
    \label{eq:Heggie's law}
\end{equation}
where $a$ is the semi-major axis of the binary, $\average{m_\ast}$ is the average mass of a star in the NSC, and $\sigma$ is the three-dimensional velocity dispersion. 
Hence we dynamically evolve hard binaries only.

\subsection{Orbital evolution and merger}
\label{sec:bbh_hardening}
Once the binary is in the migration trap, we set $R=R_\mathrm{trap}$ and update the quantities in eqs. \ref{eq:Bartos2017, Sigma gas}--\ref{eq:Bartos2017, c sound} accordingly. The semi-major axis $a$ and eccentricity $e$ of a binary with component masses $m_1$ and $m_2$, embedded in the disk, evolve due to gas hardening, irrespective of whether the binary is prograde or retrograde, as \citep{Ishibashi_2020}
\begin{align}
    \dot{a}_\mathrm{gas} &= -\frac{24 \pi \,{}\alpha\,{} c_\mathrm{s}^2 \,{}\Sigma_\mathrm{gas}\,{} (1+e)^2\,{} a}{\mu\,{} \Omega_{\rm b}} ,
    \label{eq:dot a IG} \\
\
    \dot{e}_\mathrm{gas} &= \frac{12\pi\,{}\alpha \,{}c_\mathrm{s}^2 \,{}\Sigma_\mathrm{gas}\,{}(1-e^2)^{1/2}\,{}(1+e)^2\,{}[ 1-(1-e^2)^{1/2}]}{\mu\,{} \Omega_{\rm b}\,{}e},
    \label{eq:dot e IG}
\end{align}
where $\mu=m_1 m_2 / (m_1 + m_2)$ is the reduced mass and $\Omega_{\rm b}= \sqrt{G (m_1 + m_2) /a^3}$ is the Keplerian orbital frequency. 

The binary also hardens due to the effect of GW  emission, which will govern the evolution at small semi-major axes. The evolution of the semi-major axis $\dot{a}_\mathrm{GW}$ and eccentricity $\dot{e}_\mathrm{GW}$ due to GW hardening proceeds as in \citet{Peters_1964}.

The interaction with other objects also contributes to the hardening of hard binaries according to \citet{Heggie_hard_bin}. We neglect the hardening effect due to three-body interactions because they typically occur on a timescale longer than gas hardening \citep[][see discussion in Sect.~\ref{sec:threebody}]{Leigh_McKernan_2018}.

The overall evolution of the binary is thus described by
\begin{equation}
    \dot{a}=\dot{a}_\mathrm{gas} + \dot{a}_\mathrm{GW}\,, \qquad \dot{e}=\dot{e}_\mathrm{gas} + \dot{e}_\mathrm{GW} .
    \label{eq:IG and Peters}
\end{equation}

The equations for gas hardening (eqs. \ref{eq:dot a IG} and \ref{eq:dot e IG}) are valid under the assumption that gravitational torques from the binary act axisymmetrically upon the disk so that they clear a cavity in the surrounding gas distribution, which remains circular throughout its inspiral. This is an idealized model since cavities in AGN disks can become eccentric \citep{MacFayden_2008, cimerman2023} and can lead the orbital separation to grow in time \citep{Miranda_2017_circumbinarydisk}. Hence, we will consider an `optimistic' model in which the evolution is given by 
eqs.~\ref{eq:IG and Peters} (hereafter, GH model),  and a `pessimistic' model in which we neglect gas hardening and only consider the effect of GW emission (hereafter, no-GH model). In both the optimistic and pessimistic cases, we 
integrate the hardening equations using the Euler method and an adaptive time-step 
\citep{fastcluster2021}. 

We refer to the delay time between pair-up and, eventually, merger as $t_\mathrm{del}$. If 
$t_\mathrm{pair} + t_\mathrm{del}$ 
is longer than the lifetime of the disk, which means that the disk has evaporated before the binary could merge. In this case, the BBH keeps hardening due to GW emission only. 

We assume that the BBH merges when its members cross the ISCO radius of a non-spinning BH with mass equal to the total mass of the binary system, $r_{\rm ISCO}=6\,G\,(m_1+m_2)/c^2$, 
with a tolerance of $0.1 \, r_\mathrm{ISCO}$. This happens on a merger timescale 
\begin{equation}
t_\mathrm{merg}=t_\mathrm{pair}+t_\mathrm{del} = t_\mathrm{in} + t_\mathrm{damp} + t_\mathrm{migr,\,I} +t_\mathrm{del} .
\end{equation}

We model the mass and spin of the merger
remnant using fitting formulas from numerical relativity, as described by \citet{Jimenez}. At birth, merger remnants receive a relativistic kick $v_\mathrm{kick}$ because of the transfer of linear momentum caused by asymmetries in GW emission. We use the model of \citet[eq. 14.202]{maggiore_GW} for the magnitude and direction of the kick velocity $\vec v_\mathrm{kick}$. 

The relativistic kick $\vec v_\mathrm{kick}$ pushes the merger remnant out of the migration trap. We compute the new velocity as $\vec v_\mathrm{fin}=\vec v_\mathrm{Kepl}(R_\mathrm{trap}) + \vec v_\mathrm{kick}$,
where the Keplerian velocity in the migration trap is computed accounting for the mass of the SMBH $\MSMBH$ and 
the inner part of the NSC 
$M_\mathrm{NSC}$, while neglecting the mass of the gas disk:
\begin{equation}
    v_\mathrm{Kepl} \tonde{R} = \sqrt{\frac{G \tonde{\MSMBH + M_\mathrm{NSC}(R)\,}}{R}} \equiv \sqrt{\frac{G M_\mathrm{TOT}\tonde{R}}{R}}.
    \label{eq:v_Kepler}
\end{equation}

The final semimajor axis $A_\mathrm{fin}$ of the remnant orbit, computed by means of simple orbital transfer calculations \citep{Hohmann_1925}, is
\begin{equation}
    A_\mathrm{fin}=\frac{G M_\mathrm{TOT}(R_\mathrm{trap})}{v_\mathrm{fin}^2} \; \quadre{\frac{R_\mathrm{trap}\,v_\mathrm{fin}^2 / G M_\mathrm{TOT}(R_\mathrm{trap})}{2 - R_\mathrm{trap}\,v_\mathrm{fin}^2 / G M_\mathrm{TOT}(R_\mathrm{trap})} }.
    \label{eq:R_fin}
\end{equation}
The quantities in eqs.~\ref{eq:Bartos2017, Sigma gas}--\ref{eq:Bartos2017, c sound} are updated accordingly.

As a safety check, we ensure that the new orbital semi-major axis $A_\mathrm{fin}$ is smaller than the maximum radius of the disk $R_\mathrm{max}$, 
meaning that the remnant can experience damping and become embedded in the disk. Otherwise, we 
do not consider the remnant for future generations.


\begin{figure}
    \centering
    \includegraphics[width=\linewidth]{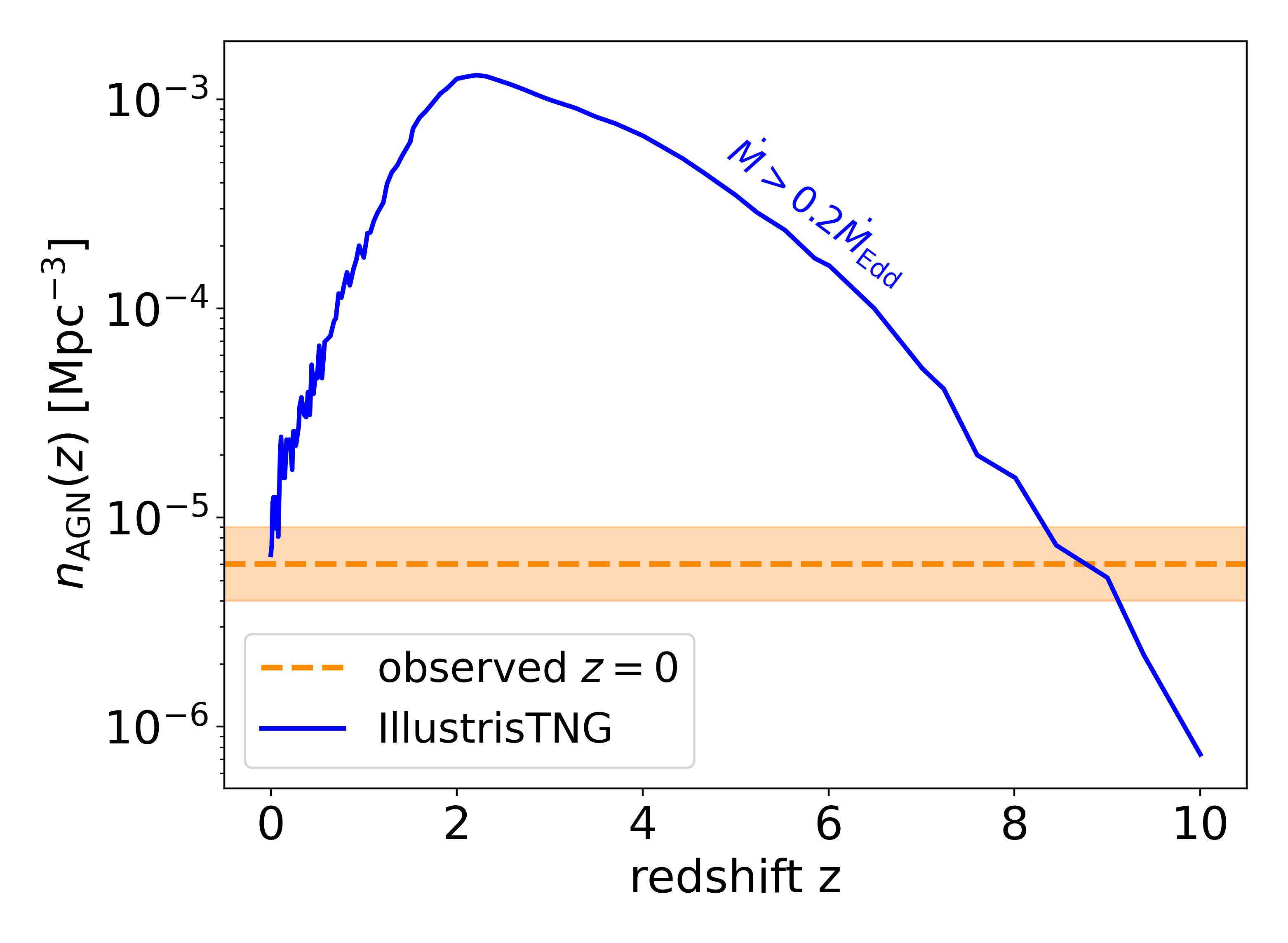}
    \caption{Number density of active SMBHs per unit comoving volume, $n_{\rm AGN}(z)$, in the {\sc IllustrisTNG100} simulation (solid blue lines) vs the observational values at redshift zero for unobscured (hydrogen column density $N_H\lesssim 10^{23}\, \si{\centi\meter^{-2}}$) high-luminosity ($L\geq 10^{43.2}$ erg s$^{-1}$) AGNs.}
    \label{fig:activenum/comovingvol}
\end{figure}

\subsection{Nth-generation (\texorpdfstring{$Ng$}{Ng})  BHs}
\label{sec:Nth_gen_model}
A seed BH can only go through a finite number of hierarchical mergers before it comes across one of the following scenarios:
\begin{enumerate}
    \item The disk has evaporated, therefore damping, migration, and any other effects due to gas torques stop. We evaluate this by checking if, at a generation $N$, the merger timescale $t_\mathrm{merg}^{(N)}$ is shorter than the disk lifetime $\tau$. In our formalism, we define the merger timescale of a $Ng$ BH as $t_\mathrm{merg}^{(N)} = t_\mathrm{in}^{(N)} + \quadre{
t_\mathrm{damp} + t_\mathrm{migr,\,I} +t_\mathrm{del}}^{(N)}$, where 
$t_\mathrm{in}^{(N)}$ 
is the evolutionary time of the previous generations.


    \item The relativistic kick received at merger is so strong that the remnant is ejected from the AGN. We compute the escape velocity considering only the gravitational potential of the SMBH and the inner NSC, while neglecting the mass of the gaseous disk, as
\begin{equation}
    v_\mathrm{esc} \tonde{R}= \sqrt{\frac{2\, G\, M_\mathrm{TOT}\tonde{R}}{R}} .
    \label{eq:escape velocity}
\end{equation}

Then we compute the final velocity after kick $\vec v_\mathrm{fin} = \vec{v}_\mathrm{in} + \vec{v}_\mathrm{kick}$ and ensure that it is smaller than the escape velocity $v_\mathrm{esc}$.
We also ensure that the merger remnant is on a disk-crossing orbit by comparing its semi-major axis $A_\mathrm{fin}$ with the outermost disk radius $R_\mathrm{max}$.

    \item The number of BHs in the NSC is finite, therefore the maximum mass that can be accreted is limited and the BH may not find any more companions to pair-up with. We keep track of the mass accreted by a single BH, that is the sum of its initial mass $m_1$ and of the masses of all the secondaries $m_2^{(N)}$ it pairs-up and merges with:
\begin{equation}
    M_\mathrm{acc}=m_1 + \sum_N m_2^{(N)} ,
\end{equation}
where the index $N$ represents the generation number.

We ensure that the BH does not accrete more mass than what is available in the inner NSC in the form of other BHs, namely $M_\mathrm{acc} \le M_\mathrm{BH}^\mathrm{max}$, 
where $M_\mathrm{BH}^\mathrm{max}$ is obtained as in eq. \ref{eq:N_BH_max}.

    \item The BH is so massive that it opens a gap in the disk and can only move from its radial location due to Type II migration. We check whether both conditions in eqs. \ref{eq:open gap} and \ref{eq:keep gap open} are respected. For typical values of viscosity and aspect ratio, these conditions entail $m_\mathrm{BH}\gtrsim 10^{-1}\MSMBH$.
\end{enumerate}

\begin{figure}
    \centering
    \includegraphics[width=\linewidth]{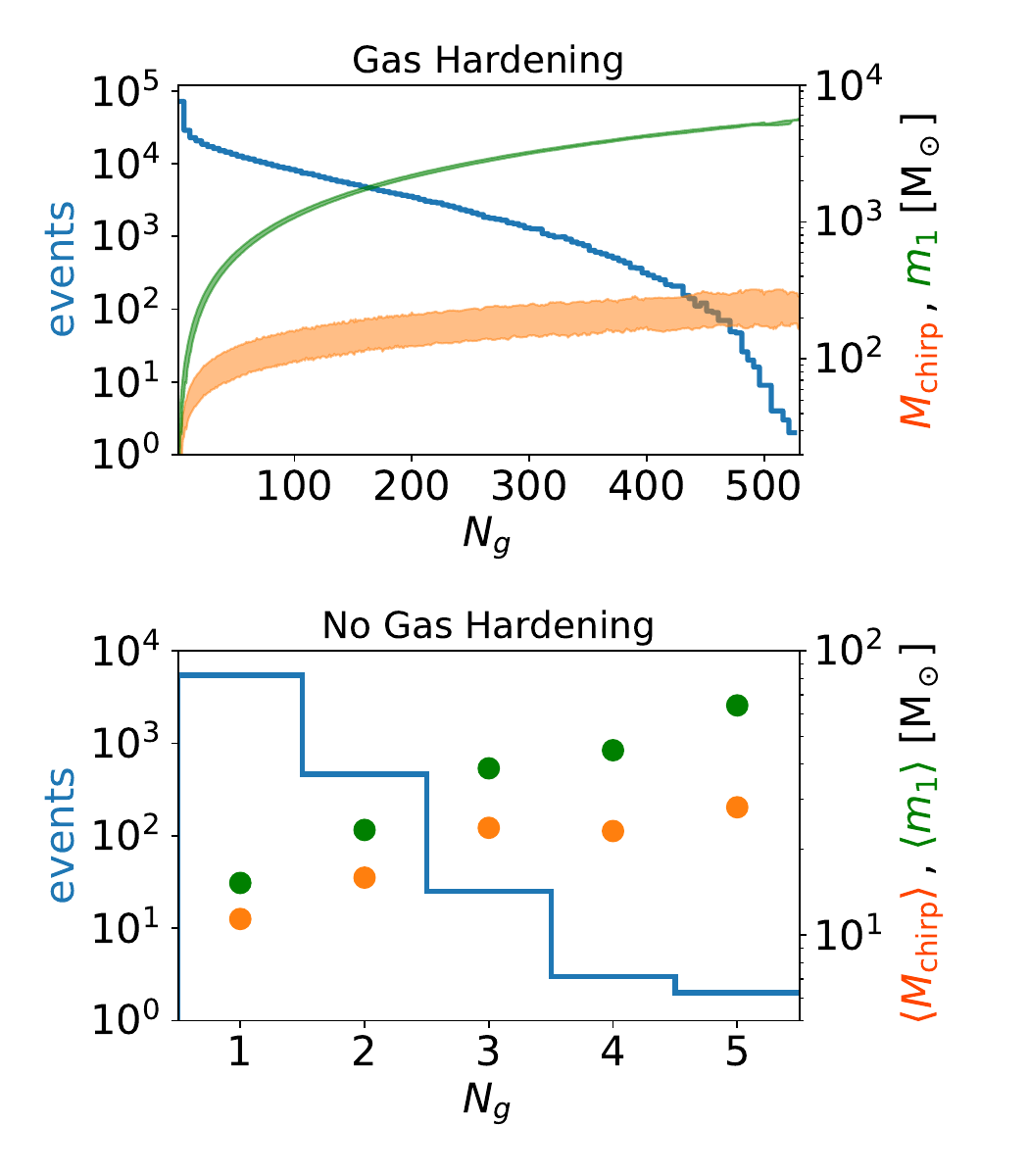}
    \caption{Blue histogram and left-hand y-axis: Number of BBH merger events for each hierarchical merger generation $N_g$. Orange (green) shaded area and upper right-hand y-axis: $25\%$ to $75\%$ percentile of the chirp mass (primary mass) for merging BBHs of each generation. Orange (green) dots and lower right-hand y-axis: average chirp mass (primary mass) for merging BBHs of each generation.}
    \label{fig:number-of-generations}
\end{figure}

\begin{figure*}
    \centering    
    \includegraphics[width=\textwidth]{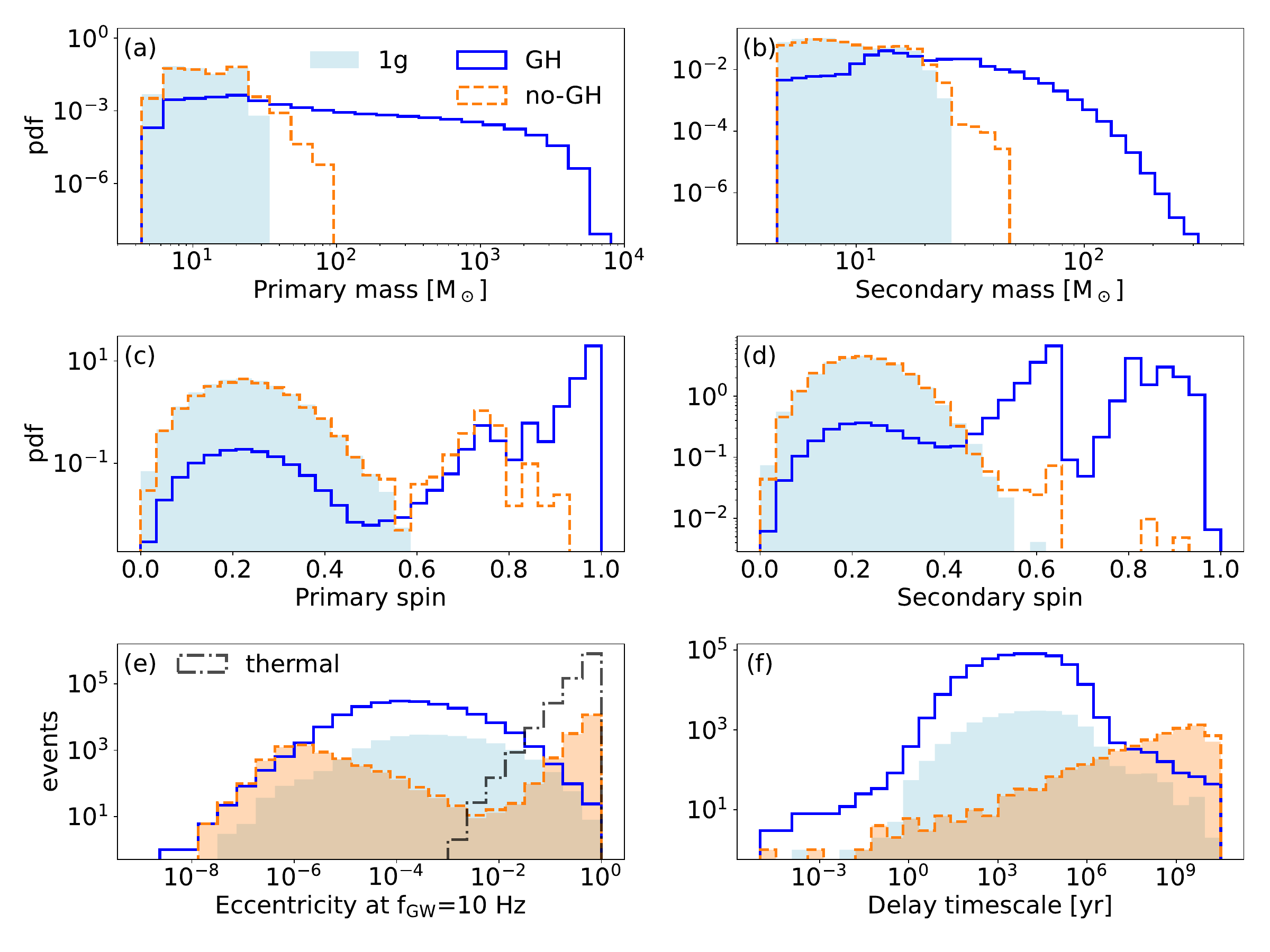}
    \caption{Main properties of dynamical BBH mergers in our gas-hardening (GH) and no gas-hardening (no-GH) AGN models. Unfilled blue solid (orange dashed) histogram: BBHs in the GH (no-GH) AGN scenario. Filled blue (orange) histogram: $1g$ BBHs in the GH (no-GH) AGN scenario. If $1g$ BHs in the no-GH scenario are identical to the GH scenario, they are not shown. \textbf{(a,b)} Probability density function (pdf) of primary and secondary BH mass ($m_1$ and $m_2$). \textbf{(c,d)} Primary and secondary spin magnitudes ($\chi_1$ and $\chi_2$). \textbf{(e)} Orbital BBH eccentricity $e(10\,{}{\rm Hz})$ when the GW frequency is $f_{\rm GW}=10$ Hz. In gray: thermal distribution $p(e)=2e$ for comparison. \textbf{(f)} Timescales of delay time $t_\mathrm{del}$ between BBH pair-up and merger.}
    \label{fig:summary}
\end{figure*}

\begin{figure}
    \centering
    \includegraphics[width=\linewidth]{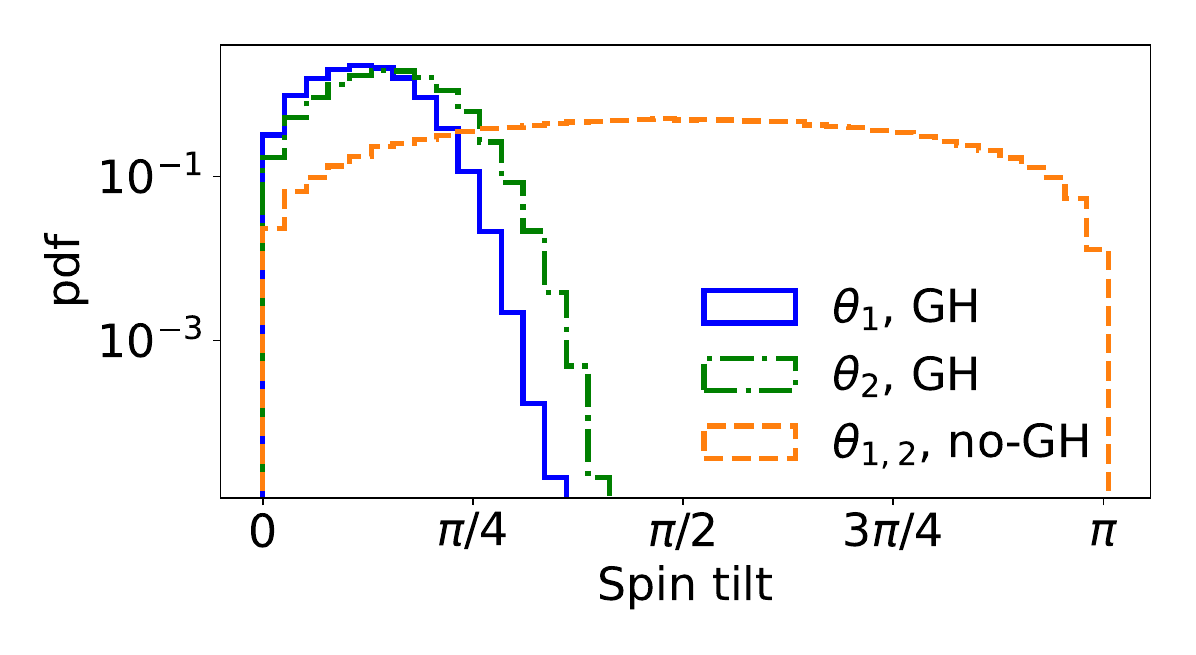}
    \caption{Probability density function (pdf) of spin tilt angles $\theta_1$ and $\theta_2$. Blue (green) histogram: primary (secondary) spin tilt in the GH scenario. Orange unfilled histogram: primary and secondary tilts in the no-GH scenario. The first generation is not shown separately in this plot because the distributions are identical at each generation.}
    \label{fig:spin-tilt}
\end{figure}

If any of the conditions 1.~--~4. is met, we 
do not consider the merger remnant for future generations. We follow the evolution of $N$th-generation BHs with a procedure similar to the one outlined for first-generation BHs. 
In hierarchical mergers, the remnant of an $(N-1)$th-generation merger acts as the primary BH for the $N$th-generation. So, the primary mass and spin are simply set as the remnant mass and spin of the previous generation, computed according to \cite{Jimenez}. 
Similarly, the initial position of the primary component of the $N$th-generation BBH is set as the position of the merger remnant of the previous generation, set in eq. \ref{eq:R_fin}. This value is used to compute the pairing time $t_\mathrm{pair}=t_\mathrm{damp} + t_\mathrm{migr,\,I} + t_\mathrm{in}$, where for $N$th-generations
$t_\mathrm{in}^{(N)}=t_\mathrm{merg}^{(N-1)}$. The migration and pair-up physics are the same as described for the first generation.

We model the secondary component of each $N$th-generation BBH as described in Appendix \ref{sec:secondary mass}. We define an $N$th-generation ($Ng$) BH to be the result of the repeated merger of $N$ stellar-origin BHs. For instance, an $Ng$ BH can either be the result of an $Mg-1g$ merger (where $M+1=N$) or of an $Mg-Lg$ merger (where $M+L=N$, $L>1$). 
We keep iterating the same procedure until the considered BH has met at least one of the requirements 1.~--~4. above.

\subsection{Cosmic evolution}
We model the cosmic evolution of AGNs using data from the {\sc IllustrisTNG100}, a  magneto-hydrodynamical cosmological simulation adopting a cubic box of size $\tonde{\SI{110.7}{\mega \parsec}}^3$ with a resolution of roughly $\tonde{\SI{6}{\kilo\parsec}}^3$ \citep[see,][for further details]{TNG100_1, TNG100_2}.

We calibrate the AGN density distribution $n_{\rm AGN}(z)$ in the {\sc IllustrisTNG} data by choosing a threshold in the mass accretion rate such that the value at redshift zero, $n_{\rm AGN}(0)$, is compatible with the observational value of $\sim\SI{6e-6}{\mega\parsec^{-3}}$ obtained from a sample of X-ray-selected AGNs \citep{Buchner_2015}. As displayed in Figure~\ref{fig:activenum/comovingvol}, we find that setting this threshold to a fifth of the Eddington mass accretion rate, namely $\dot{M}_\mathrm{SMBH} \geq 0.2\, \dot{M}_\mathrm{Edd}$, provides an appropriate normalization.
We employ $n_{\rm AGN}\tonde{z}$ to compute the BBH merger rate properties in various redshift bins as explained in Appendix~\ref{sec:cosmorate}.

The SMBH mass distribution is expected to evolve as a function of redshift \citep{IllustrisTNG_SMBH}. Nevertheless, since the mass evolution 
cannot be constrained by current data, we use the observational distribution from \cite{Greene_2007} at every redshift for simplicity.

\begin{figure*}
    \centering    
    \includegraphics[width=\textwidth]{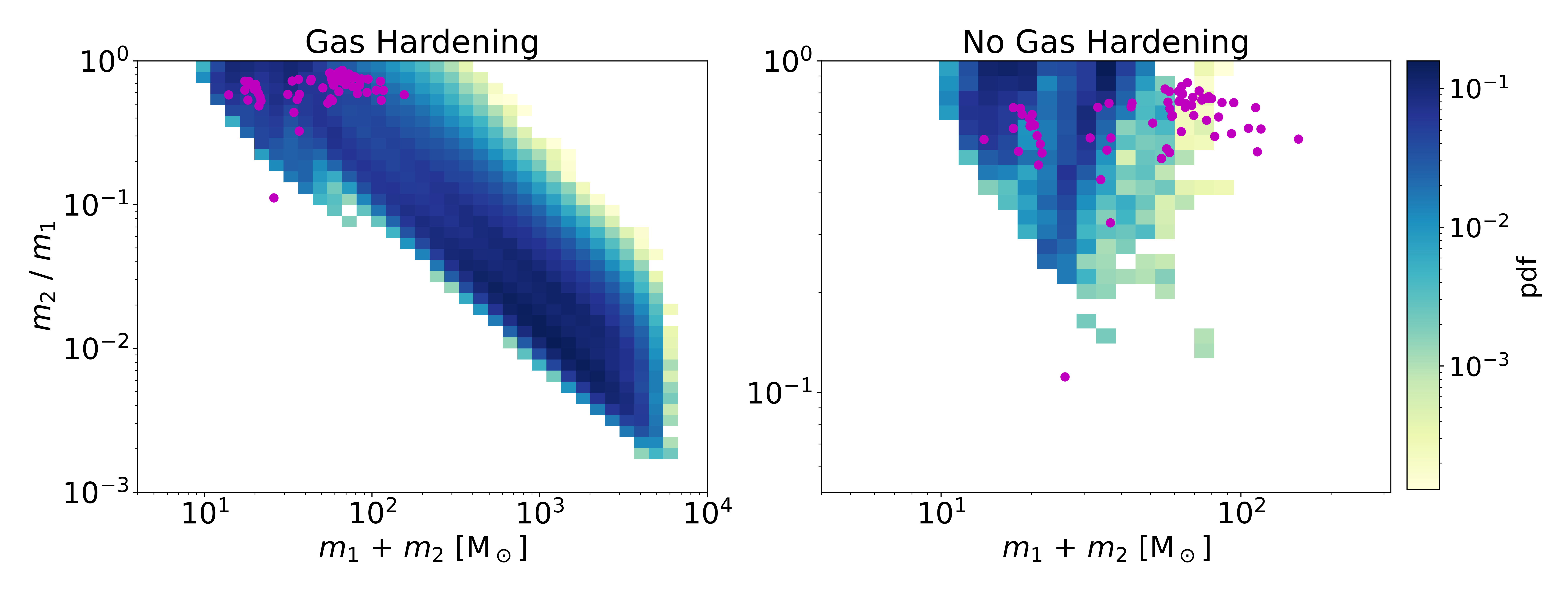}
    \caption{Probability density function (pdf) of binary mass and mass ratio for all BBH mergers in our simulations (blue color palette). We do not apply any selection effects to our simulations. Left-hand (right-hand) panels: GH (non-GH) scenario. The magenta dots show the median values of the parameters for Ligo-Virgo-KAGRA (LVK) BBH merger event candidates with $p_\mathrm{astro}>0.9$ from GWTC-3 \protect\citep{GWTC3_first, GWTC3_second}. We do not include error bars for readability purposes.} 
    \label{fig:mass-ratio}
\end{figure*}

\begin{figure*}
    \centering    
    \includegraphics[width=\textwidth]{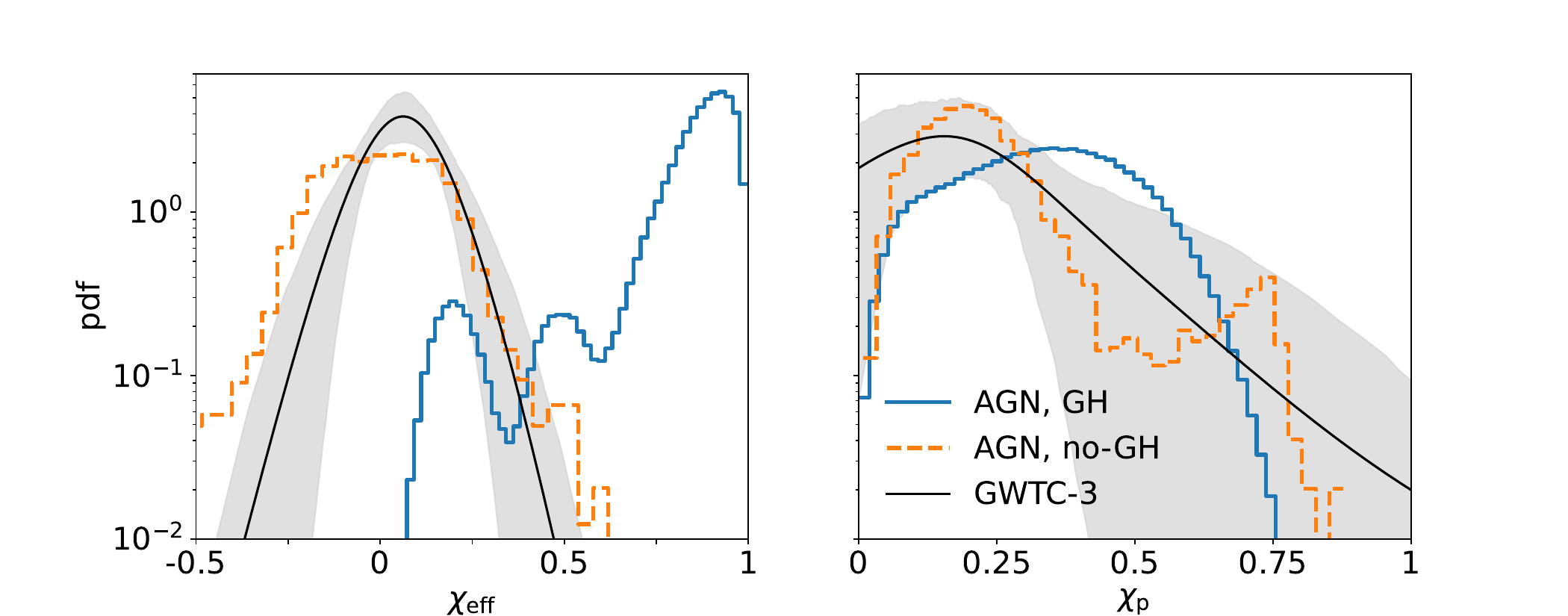}
    \caption{Effective (left) and precession spin (right) probability density functions (pdf) for the gas-hardening (GH, solid blue lines) and  non-gas-hardening 
    (no-GH, dashed orange lines) model. We do not apply any selection effects to our simulations.
    Black solid lines: posterior distribution inferred from LVK data, after applying a parametric-model description of the intrinsic BBH population with hierarchical Bayesian inference \protect\citep{GWTC3_second}.
    }
    \label{fig:spin-LVK}
\end{figure*}

\begin{figure*}
    \centering    
    \includegraphics[width=0.9\textwidth]{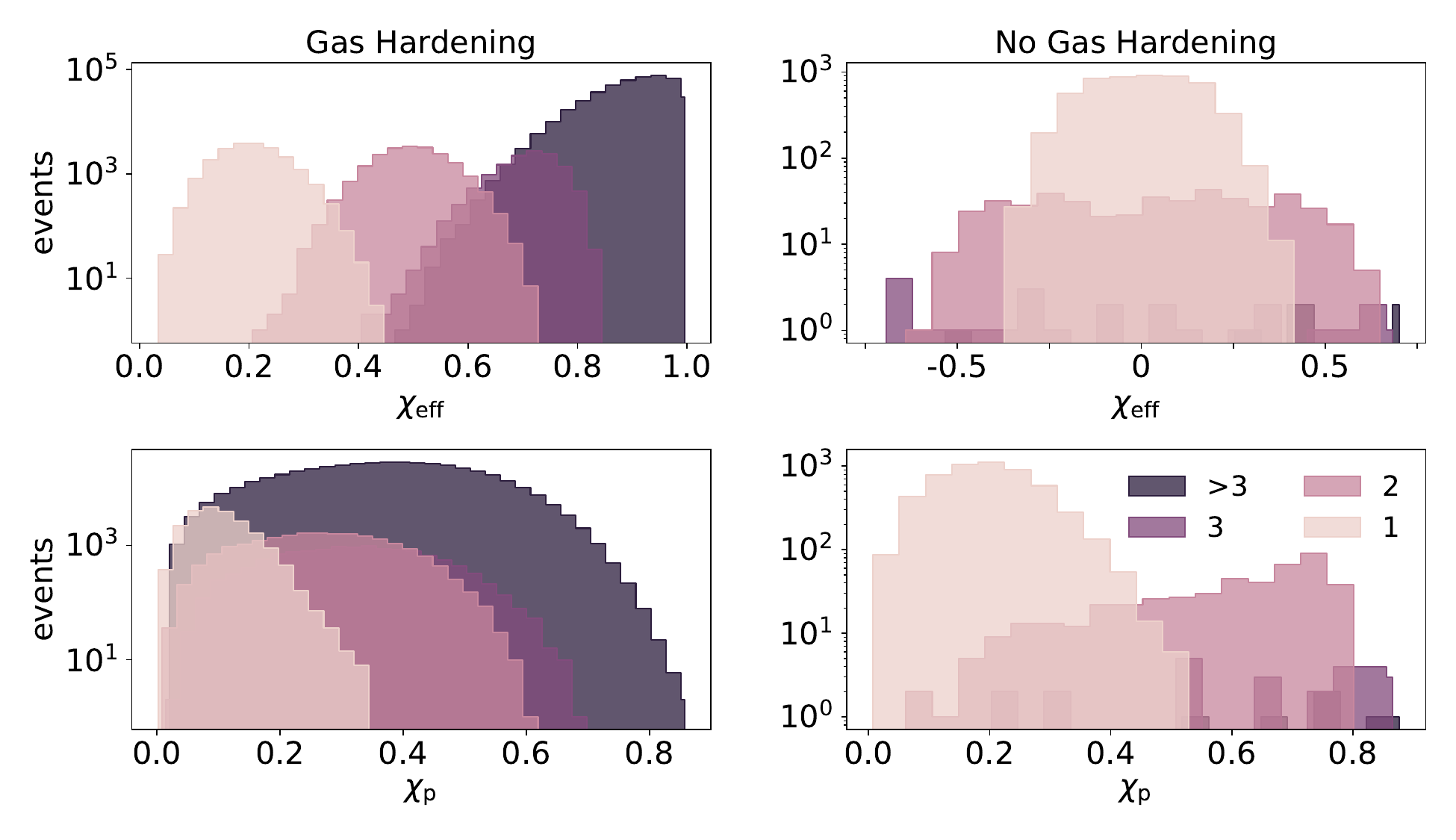}
    \caption{Effective and precession spin distributions for the GH (left-hand panels) and no-GH (right-hand panels) models, where the shade represents different hierarchical merger generations. The lightest hue refers to the first generation (stellar-origin BHs), while the darkest hue refers to generations higher than the third.}
    \label{fig:ge-eff-prec-spin}
\end{figure*}

\subsection{Description of runs}

We consider two different models for the AGN channel: with and without gas hardening (hereafter, GH and no-GH). We have run $N=10^5$ realizations of our two models,
each with different SMBH mass and AGN lifetime randomly extracted as described at the end of Section~\ref{sec:setup_AGN_methods}. In each run, we simulate all the BHs that reach the migration trap within the AGN disk lifetime. 

Furthermore, we simulate four other channels (isolated, YSC, GC, NSC) with the same code (\fastcluster{}, \citealt{fastcluster2021,fastcluster2022}) and using the same underlying initial conditions, such as the stellar evolution model determining the $1g$ BHs mass distribution (Appendix~\ref{sec:oldfastcluster}). This allows us to filter out any bias that might arise by using different numerical codes  for different environments: the differences we see are not due to the numerical approach adopted but to 
intrinsic 
differences among channels.  We perform a Bayesian analysis to determine the AGN mixing fraction as described in Appendix~\ref{sec:mix_frac}.

\section{Results}
\label{sec:results}


\subsection{Impact of gas hardening}
We find that the efficiency of the hierarchical merger process is significantly higher in the presence of gas hardening: in our gas-hardening (GH) model, seed BHs can go through up to roughly 500 merger episodes, whereas in the no gas-hardening (no-GH) case the hierarchical chain usually stops after a few generations (Fig.~\ref{fig:number-of-generations}). In the GH (no-GH) model, the hierarchical merger chain  stops because of the condition on timescales, maximum mass, Type II migration, or ejection in the $58.7\%$ ($81.6\%$), $27.1\%$ ($18.0\%$), $13.6\%$ ($0\%$), and $0.6\%$ ($0.4\%$) of simulations, respectively. 

Figure~\ref{fig:summary} shows the main properties of dynamically assembled BBHs in our AGN simulations. We display only BBHs that merge within a Hubble time. 
In both models, hierarchical mergers in the migration trap are successful in producing BBHs with primary mass 
in the pair-instability mass gap and above, but there is a major difference between the GH and no-GH scenarios: the primary BH mass extends only up to $300\, \Msun$ in the no-GH scenario, while it reaches $\approx{5}\times{}10^3 \Msun$ in the GH scenario, because gas hardening dramatically increases the efficiency of hierarchical mergers.
Hence, the gas hardening mechanism produces a recognizable feature in the high-mass end ($m_1 \gtrsim 100 \Msun$) of the mass spectrum.

This is caused by the stark difference in delay timescales (Fig.~\ref{fig:summary}f) due to the different gas hardening prescriptions: in the GH scenario, the relative distribution has a pronounced peak between $1\,{\rm yr}$ and $1\,{\rm Myr}$, which in the no-GH case is absent. 
Information on other relevant timescales is illustrated in Section~\ref{sec:timescales}.

Secondary BHs can also have masses in the pair-instability mass gap but only up to a few~$\times{}10^2 \Msun$ in the GH scenario, as the AGN channel tends to favor mergers with low mass ratio (Fig.~\ref{fig:summary}b). 

The GH model has a sharp peak in the distribution of the primary spin magnitude at $\chi_{1}=1$ as well as a peak in the distribution of the secondary spin magnitude at $\chi_{2}\simeq 0.9$ associated with high-generation mergers (Fig.~\ref{fig:summary}c,d). Both models GH and no-GH also have a peak at $\chi_{1}\approx{0.75}$ (also $\chi_{2}\approx{0.75}$ for the GH case), corresponding to the second generation. 

The distribution of spin-tilt angles is influenced by gas hardening, as shown in Fig.~\ref{fig:spin-tilt} and explained in Appendix~\ref{sec:spin_tilt}. In the no-GH scenario, 
the spin tilts $\theta_{1,2}$  are isotropically distributed with respect to the orbital angular momentum $\vec L$. 
Instead, in the presence of gas hardening, 
$\vec\chi_1$, $\vec\chi_2$ and $\vec L$ are all preferentially aligned with 
each other. The distribution of the secondary spin tilt $\theta_2$ is slightly wider than the primary's spin tilt, because of a cumulative effect: the misalignment between $\vec\chi_2$ and $\vec L$ is set keeping into account the misalignment between $\vec\chi_1$ and $\vec L$ (see Appendix~\ref{sec:spin_tilt}).  

Both scenarios preferentially produce BBHs with low eccentricity when\footnote{Here we only refer to BBH mergers with ISCO frequency in the LVK detectability range, $f^\mathrm{ISCO}_\mathrm{GW}\geq\SI{10}{\hertz}$. See Section~\ref{sec:ETandLISA} for more details.} the GW frequency is $\SI{10}{\hertz}$. Nevertheless, they can produce BBHs with eccentricity $e\geq 10^{-2}$ (Fig.~\ref{fig:summary}e), which is potentially detectable by the LVK interferometers \citep{Romero_Shaw_2021}. In our GH model, there are two processes at play: gas hardening pumps the eccentricity (eq.~\ref{eq:dot e IG}), whereas GW emission damps it \citep{Peters_1964}. Hence, the GH scenario is more likely to produce BBHs with eccentricity in the range $e\in[10^{-5},10^{-2}]$ for $f_\mathrm{GW}=\SI{10}{\hertz}$ than the no-GH one.  
In a more accurate model, we should also include the effect of three-body scatterings \citep{Samsing_2022} and tidal forces exerted by the SMBH \citep{Rom_2023}, both of which pump BBH eccentricity.

Figure~\ref{fig:mass-ratio} shows the relation between the BBH mass $m_1+m_2$ and mass ratio $m_2/m_1$ in our models, compared to  GW data \citep{GWTC3_first, GWTC3_second}.
The AGN channel tends to favor mergers with low mass ratio (Appendix~\ref{sec:secondary mass}). 
This effect is particularly 
important for high primary BH masses $m_1\gtrsim10^2\Msun$ when accounting for gas hardening. 

The effective spin distribution in the GH model (Fig.~\ref{fig:spin-LVK})
peaks at $\chi_{\rm eff}\sim{1}$, corresponding to maximum alignment, and shows lower peaks at $\chi_\mathrm{eff}\simeq0.2$ and $\chi_\mathrm{eff}\simeq 0.5$. From Fig.~\ref{fig:ge-eff-prec-spin}, 
we see that $1g$ mergers populate the peak at $\chi_\mathrm{eff}\simeq0.2$, $2g$ mergers create a peak at $\chi_\mathrm{eff}\simeq0.5$, while third- and higher-generation BBHs contribute the main peak at $\chi_\mathrm{eff}\simeq1$. The precession spin distribution, instead, has no sharp features in this scenario.

In the no-GH scenario, the orbital angular momentum is isotropically oriented with respect to the AGN disk (Appendix~\ref{sec:spin_tilt}). This produces a bell-shaped effective spin distribution centered on $\chi_\mathrm{eff}=0$ with half-width at half-maximum $\mathrm{HWHM}\simeq0.3$ and a precession spin distribution that peaks at $\chi_\mathrm{p}\simeq0.2$ and $\chi_\mathrm{p}\simeq0.75$ (Fig.~\ref{fig:spin-LVK}). As the BBH generation increases, so does the magnitude of its BH spins, making the corresponding effective spin distribution 
progressively wider as shown in Fig.~\ref{fig:ge-eff-prec-spin}. The peaks at $\chi_\mathrm{p}\simeq0.2$ and $0.7$ are populated mostly by $1g$ and $2g$ mergers, respectively, whereas  the tail at $\chi_\mathrm{p}>0.8$ results from higher-generation mergers. 

\begin{figure*}
    \centering
     \includegraphics[width=\textwidth]{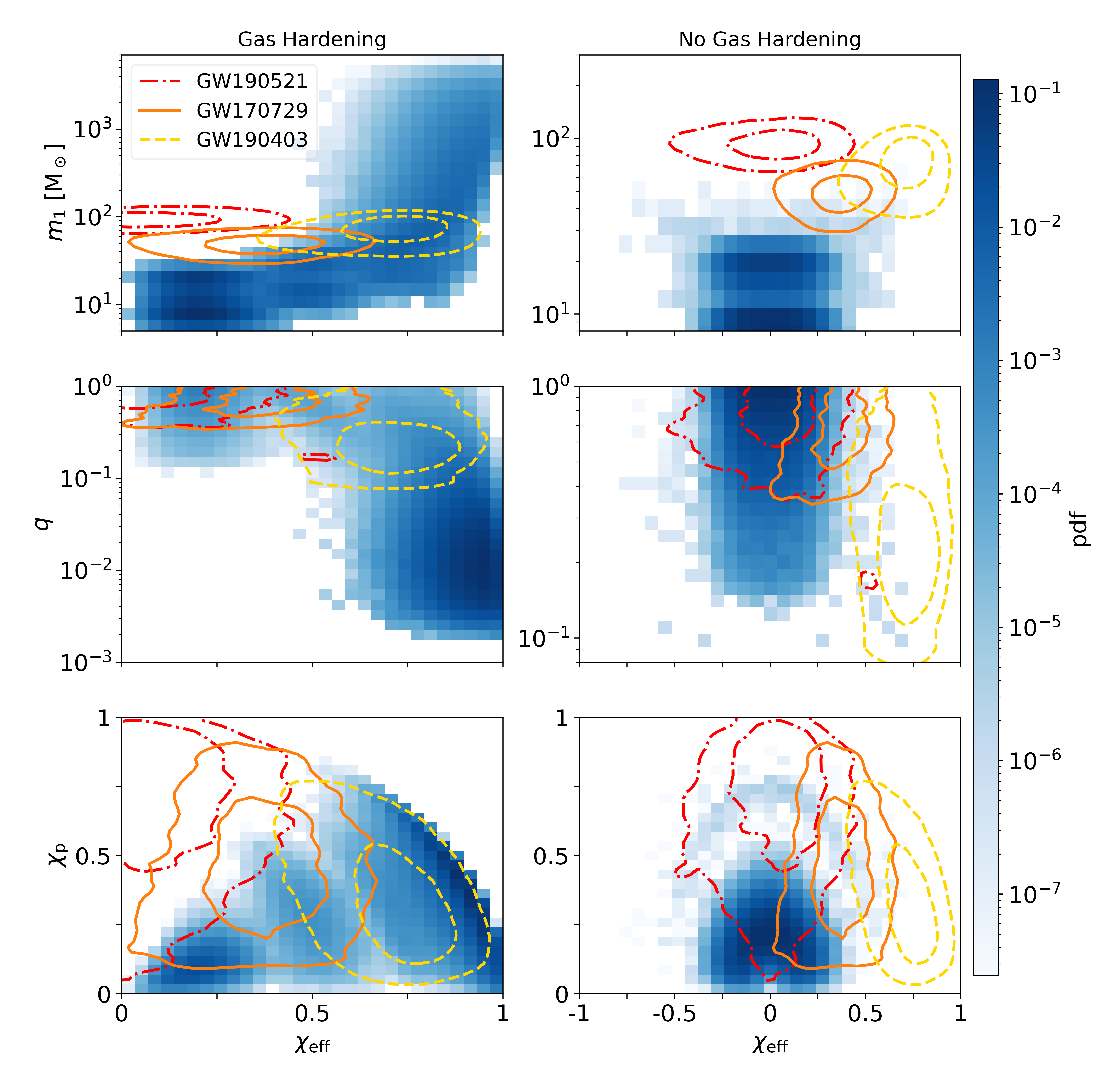}
    \caption{Top row: probability density distribution for primary BH mass $m_1$ and effective spin $\chi_\mathrm{eff}$ compared with posterior contour plots at credibility levels $50\%$ and $90\%$ of LVK BBH merger events GW190521 (red dash-dotted line, \citealt{GW190521}) and GW170729 (orange solid line, \citealt{GWTC3_second}), and LVK transient event GW190403$\_$051519 (yellow dashed line, \citealt{GWTC2.1}). Central (bottom) row: Same as the first row but for mass ratio $q=m_2/m_1$ (precession spin $\chi_\mathrm{p}$) and effective spin $\chi_\mathrm{eff}$.}
    \label{fig:hist-LVK-comparison}
\end{figure*}

\subsection{Anti-correlation between \texorpdfstring{$q$}{q} and \texorpdfstring{$\chi_{\rm eff}$}{Xeff}}
Figure~\ref{fig:hist-LVK-comparison} shows the relationship of the effective spin with primary mass and mass ratio of BBH mergers in our simulations. In the GH scenario, there is a clear correlation between effective spin and BBH mass, as well as a clear anti-correlation between effective spin $\chi_{\rm eff}$ and mass ratio $q$. This is a possible explanation for the anti-correlation between $\chi_{\rm eff}$ and $q$ found in the LVK data \citep{Callister_2021}. If this anti-correlation stems from hierarchical mergers in the gas-hardening regime, our simulations suggest that it should extend to lower mass ratios and higher effective spins than currently observed  
by LIGO and Virgo. 

The anti-correlation is particularly noticeable for $\chi_\mathrm{eff}\gtrsim 0.4$ because all higher-generation mergers display the following features: high mass, small mass ratio, and high effective spin. 
In the no-GH model, both the spin alignment and the hierarchical merger process are suppressed; so there is no clear correlation between the aforementioned quantities.

\begin{figure*}
    \centering
    \includegraphics[width=\textwidth]{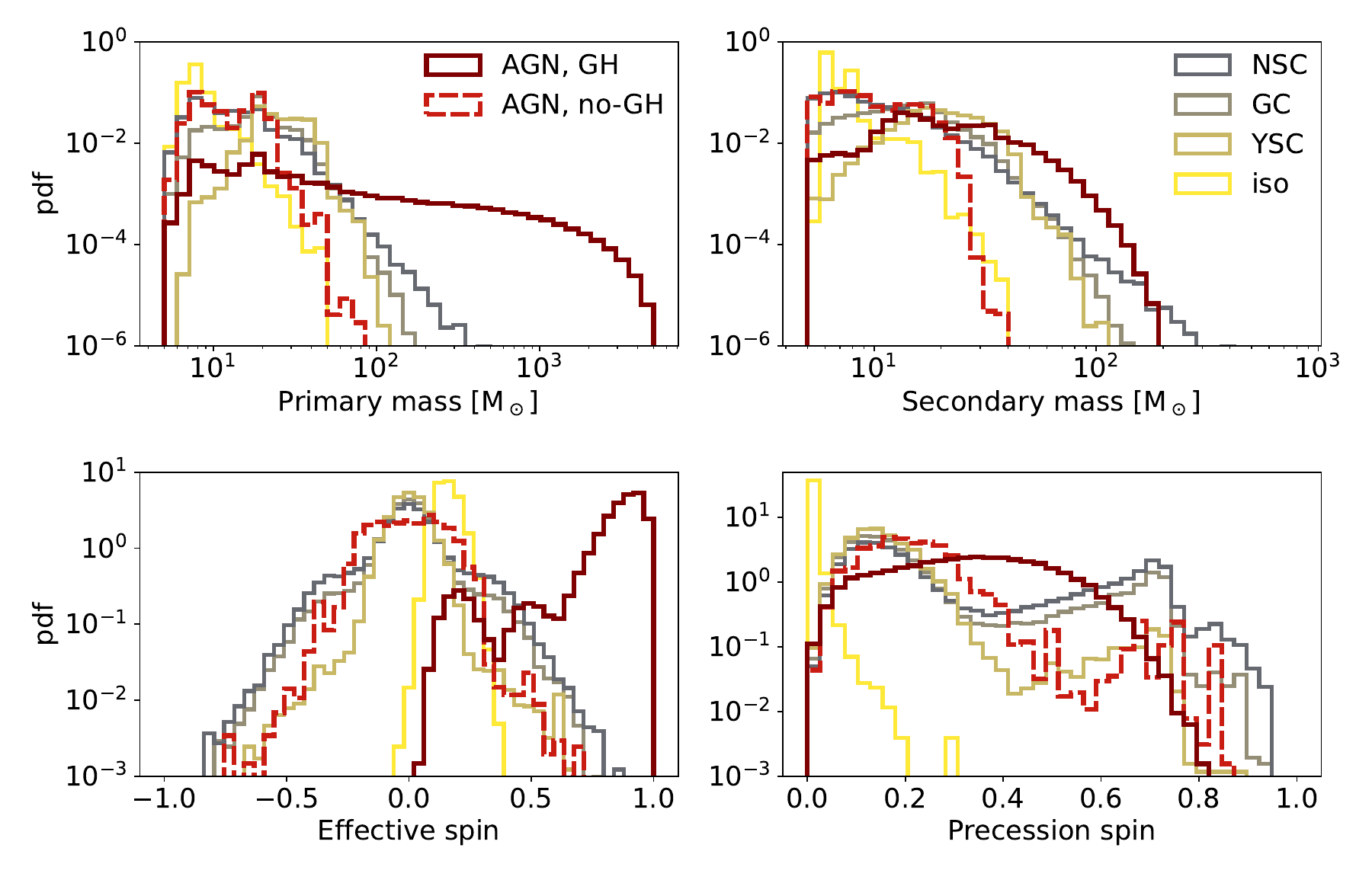}  
    \caption{Probability density distribution (pdf) of primary and secondary BH masses $m_{1,2}$, effective spin $\chi_\mathrm{eff}$ and precession spin $\chi_\mathrm{p}$ in the five channels modeled with \fastcluster{}: active galactic nuclei (AGNs) with gas hardening (GH, dark red solid line) or without (no-GH, light red dashed line), nuclear star clusters (NSCs, dark gray line), globular clusters (GCs, light gray line), young star clusters (YSCs, dark gold line), and isolated binary evolution (iso, light gold line).} 
    \label{fig:multichannel}
\end{figure*}

\begin{figure*}
    \centering
    \includegraphics[width=0.45\linewidth]{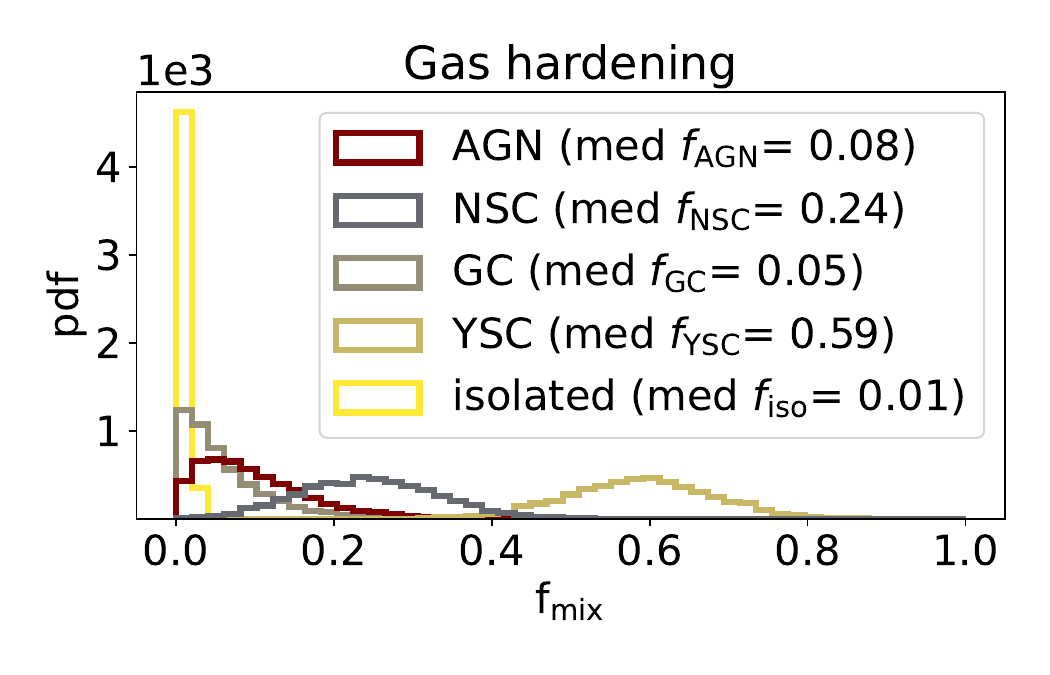}
    \hspace{1cm}
    \includegraphics[width=0.45\linewidth]{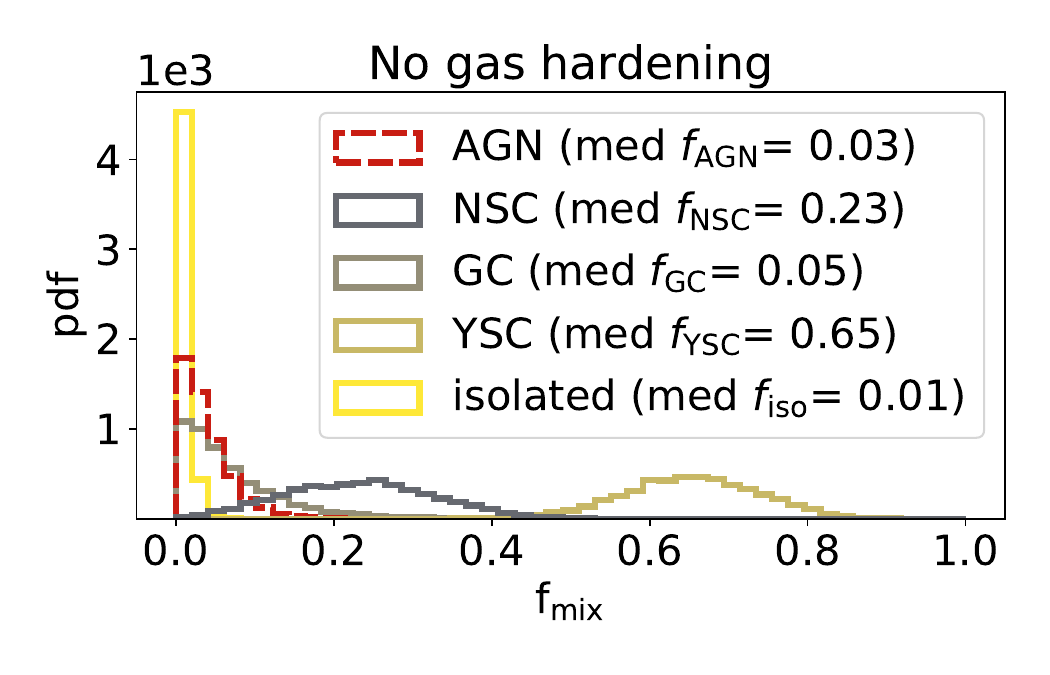}
    \caption{Mixing fractions obtained with our five-channel analysis (see Appendix~\ref{sec:mix_frac}). Left-hand (right-hand) plot: GH (no-GH) AGN channel model.}
    \label{fig:mixing-fraction}
\end{figure*}

\subsection{Comparison with other channels}
\label{sec:multichannel}
Figure~\ref{fig:multichannel} 
compares the main properties of BBH mergers in five different formation channels, namely AGN disks, NSCs, GCs, YSCs, and isolated binary evolution (iso).  
The simulations for NSCs, GCs, YSCs, and iso adopt the same set-up as model B of \cite{fastcluster2022}, and use as initial conditions catalogs of BH masses derived with {\sc sevn} \citep{sevn_2023}, for consistency with the catalogs of the AGN channel. We provide more details about  NSC, GC, YSC,  and iso simulations in Appendix~\ref{sec:oldfastcluster}.

All dynamical formation channels can produce BBH mergers with masses and spin magnitudes higher than the isolated channel. The no-GH AGN disk scenario resembles most closely the results of the other dynamical channels, whereas the GH AGN disk channel is completely different from the others. 

The population of dynamical channels appears as follows: the mass distributions show one or two peaks at a few ten solar masses and a tail extending up to a few hundred solar masses, 
the effective spin distribution is symmetric, peaks at $\chi_\mathrm{eff}= 0$ and has an HWHM of $\sim 0.5$ ($\sim 0.3$ for the YSC channel), 
whereas the precession spin distribution has two peaks at $\chi_\mathrm{p}\simeq 0.2$ and $\chi_\mathrm{p}\simeq 0.7$. This contrasts with the isolated channel population which features a primary (secondary) BH  mass distribution that does not extend any higher than $50 \Msun$ ($40 \Msun$).

The primary BH mass distribution for the GH AGN channel, instead, 
extends up to 
$\sim \SI{5e3}{\Msun}$. 
The effective spin distribution has two sharp peaks at $\chi_\mathrm{eff}\simeq 0.2$ and $\chi_\mathrm{eff}\simeq 1$, as well as a minor peak at $\chi_\mathrm{eff}\simeq 0.5$, whereas the precession spin distribution has no evident peaks.

We compute the mixing fractions $f_{\rm mix}$ for our five channels, as described in Appendix~\ref{sec:mix_frac}. To derive $f_{\rm mix}$, we have marginalized over the merger rate density, to avoid that this extremely uncertain quantity affects our results. As shown in Fig.~\ref{fig:mixing-fraction}, the NSC and YSC channels are associated with larger median mixing fractions than the other channels. This happens because, in our models, the NSC channel can account for the peak at low BH mass ($m_1\approx{10}$~M$_\odot$) found in the LVK data, while the YSC channel contributes mostly to the high-mass peak at $\sim{35}$~M$_\odot$ \citep{GWTC3_second}. In fact, in our model, the YSC scenario produces larger median BBH masses than the NSC scenario, because the escape velocity from YSCs is much lower, preventing the low-mass BHs from merging in YSCs (they are ejected by supernova kicks, see the discussion in \citealt{fastcluster2021}). The isolated channel also produces BBH mergers with a peak of the primary BH mass at $\sim{10}$~M$_\odot$, but 
has too little support for zero and negative values of $\chi_{\rm eff}$  
with respect to the observed ones.  The mass and spin properties of the GC scenario are intermediate between YSCs and NSCs. 

There is a small  difference between the mixing-fraction distribution of the GH and no-GH scenarios because this analysis only considers detectable events: the most massive BHs of the GH population (which are the main difference with respect to the no-GH model) have no impact on $f_{\rm mix}$.  Overall, our results confirm that the current LVK sample of BBH mergers is too small and the uncertainties on theoretical models are too large to draw an informative mixing-fraction analysis with five channels (see Section~3.4 of \citealt{fastcluster2022}). 

\subsection{Timescales}
\label{sec:timescales}

\begin{figure*}
    \centering
    \includegraphics[width=\textwidth]{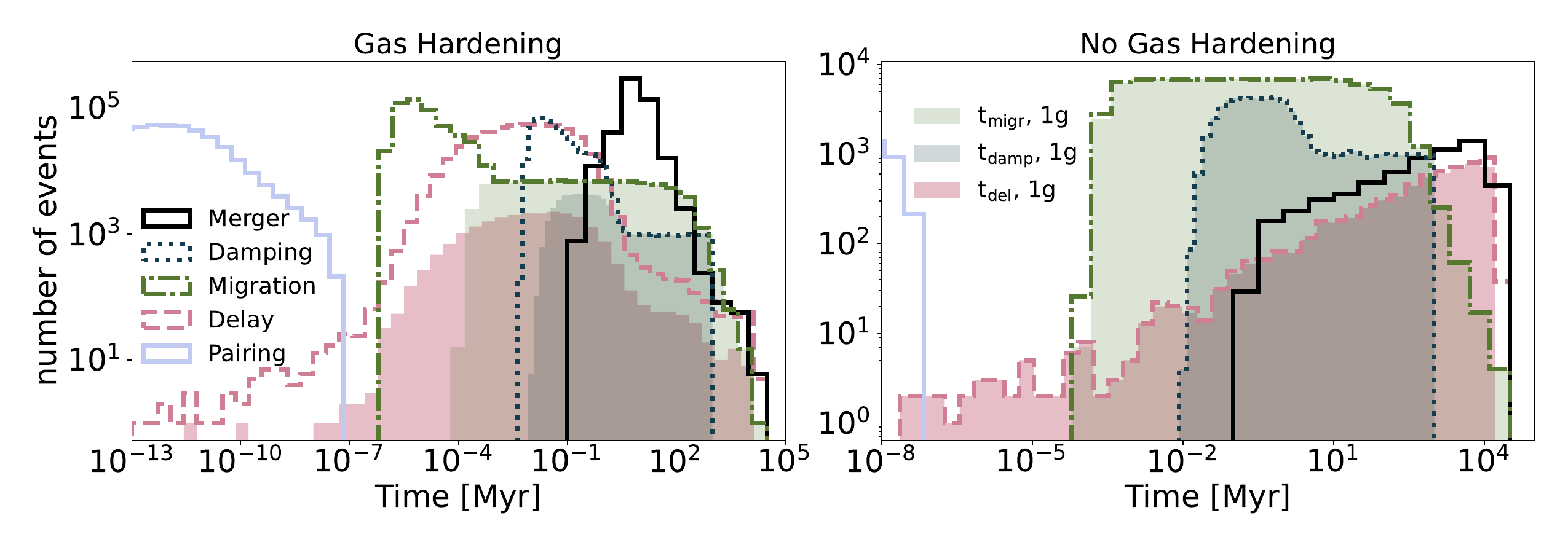}
    \caption{Histograms of the relevant timescales in our model. Black line: overall merger timescale $t_\mathrm{merg}$. Green dash-dotted line: gas capture timescale $t_\mathrm{damp}$ (the green shaded histogram shows $t_\mathrm{damp}$ for $1g$ BBHs only). Navy dotted line: Type I migration timescale $t_\mathrm{migr,\,I}$ (the navy shaded histogram  shows $t_\mathrm{migr,\,I}$ for $1g$ BBHs only). Pink dashed line: delay timescale $t_\mathrm{del}$ between pair-up and merger (the pink shaded histogram shows $t_\mathrm{del}$ for $1g$ BBHs only). Light blue line:  gas dynamical friction timescale $\tau_0$ related to BBH pairing \protect\citep{Qian_2023}.}
    \label{fig:timescales}
\end{figure*}

\autoref{fig:timescales} shows the distribution of all relevant timescales with and without gas hardening.
Both the damping time $t_\mathrm{damp}$ and the migration time $t_\mathrm{migr,\,I}$ span a large range, which is representative of the large range in SMBH mass and initial position $R$.
The damping timescale for the first hierarchical merger generation has a flat distribution in the range $10^{-4} - 10^{2}\, \si{\mega\year}$, while the migration timescale ranges between $10^{-2}$ and $10^{3}\, \si{\mega\year}$. 
In the GH scenario, $t_\mathrm{damp}$ and $t_\mathrm{migr,\,I}$ display additional peaks for subsequent generations respectively at $10^{-5}\,\si{\mega\year}$ and $10^{-2}\,\si{\mega\year}$. 
The delay timescale $t_{\rm del}$ 
spans a large range of values, because of the large range encompassed in BH masses and initial BBH semi-major axes. In the GH scenario, it has a pronounced peak at $\sim\SI{1}{\year}$, whereas in the no-GH case it is typically as large as the Hubble timescale ($\sim\SI{14}{\giga\year}$). Indeed, the longest timescale in the no-GH scenario is the delay timescale. 
In the GH scenario, instead, the evolution is overall governed by the migration timescale.

We also display the timescale for gas dynamical friction $\tau_0$, responsible for BBH pair-up, 
which we discuss in Section~\ref{sec:discussion_BBH_pairing}. It is typically of the order of $10^{-13}\,\si{\mega\year}$ and can be as short as $10^{-20}\,\si{\mega\year}$ for the disk parameters in our model; hence, it is negligible compared to other timescales at play.

\section{Discussion}
\label{sec:discussion}

\begin{figure}
    \centering
    \includegraphics[width=\linewidth]{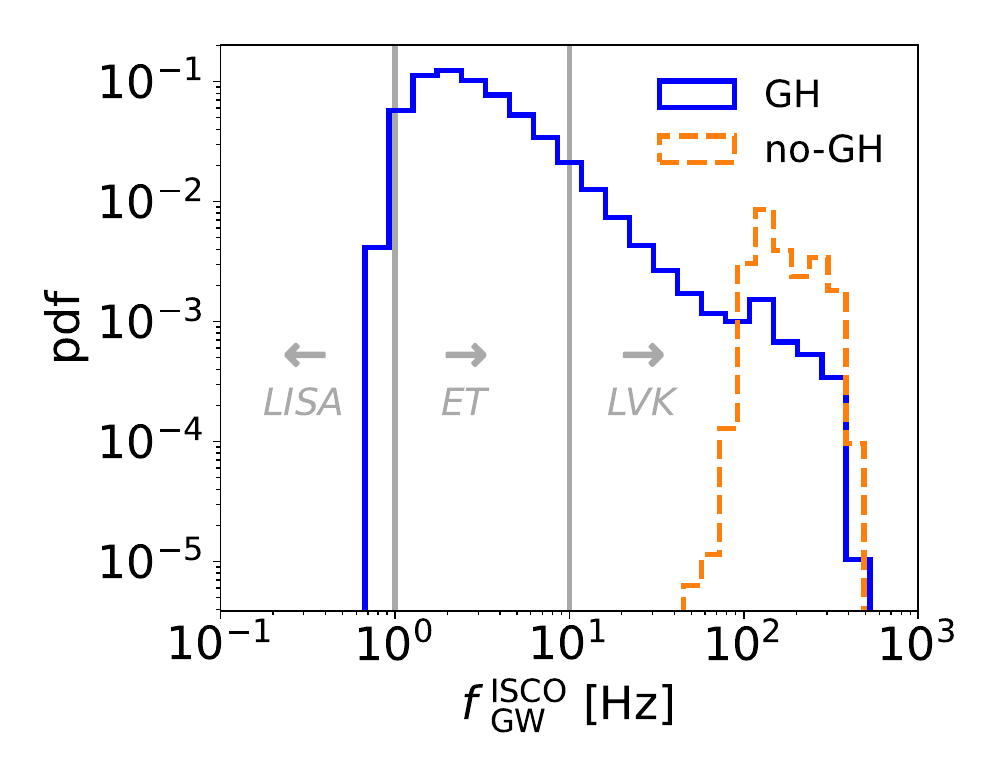}
    \caption{Solid blue (dashed orange) histograms: probability density functions (pdf) for ISCO emitted GW frequencies in the GH (no-GH) scenario. Gray lines: 
    approximate detectable frequencies of LVK detectors, 
    Einstein Telescope (ET),  
    and Laser Interferometer Space Antenna (LISA).}
    \label{fig:predictions_ET_LISA}
\end{figure}

\subsection{Outlook for ET and LISA} 
\label{sec:ETandLISA}
We have studied the evolution of hierarchical mergers in the migration traps of AGN disks and explored the role of gas hardening. Our models, especially 
the gas-hardening (GH) scenario, predict the formation of merger events with higher mass than is detectable by current ground-based detectors. 
Indeed, the frequency emitted by a BBH  depends on its mass and steadily increases during its inspiral. The frequency of the ISCO 
is equal to \citep{maggiore1_GW}
\begin{equation}
    f_\mathrm{GW}^\mathrm{ISCO}=\frac{1}{\pi\,{} 6 \sqrt{6}}\, \frac{c^3}{G\tonde{m_1+m_2}} \simeq \SI{4.4}{\kilo \hertz} \tonde{\frac{\Msun}{m_1+m_2}} .
    \label{eq:f_gw_max}
\end{equation}

Figure~\ref{fig:predictions_ET_LISA} shows the probability distribution function of the maximum emitted frequency by AGN BBHs. 
The LVK interferometers are only sensitive in the frequency range from a few ten Hz up to $\sim \SI{1}{\kilo \hertz}$ \citep{GWTC3_first}, 
hence only a fraction of our predicted merger events in the GH scenario are observable with existing detectors.

The next generation ground-based (Einstein Telescope and Cosmic Explorer, \citealt{Punturo_2010_ET,Maggiore_2020_ET,evans2021_CE,Branchesi_2023_ET}) and space-borne interferometers (LISA, DECIGO, and TianQin, \citealt{Amaro_Seaone_2013_LISA, Amaro_Seaone_2017_LISA, Luo_2016_TianQuin, Kawamamura_2019_decigo}) will be able to detect GW signals with lower frequency than currently possible. For example, the frequency range of the Einstein Telescope and Cosmic Explorer ($\approx{1-10^4}$~Hz) 
has substantial overlap with the GW frequency of BBH mergers from the GH AGN channel, with the exception of the very high-mass tail. 
LISA, instead, will be sensitive to frequencies lower than $\sim\SI{1}{\hertz}$ \citep{LISA_sensitivity_2019}, so it would be able to detect the highest-mass end of the synthetic mergers predicted by our GH AGN model. We will explore detectability by ET and LISA in detail in a follow-up work.

\subsection{BBH pair-up}
\label{sec:discussion_BBH_pairing}
In our model, we assume that the pairing of a primary and a secondary BH is immediate as soon as the primary reaches the migration trap. A realistic model for BBH pair-up in AGN disks should keep into account GW two-body capture, three-body encounters, and gas dissipation. \citet{whitehead_2023} run hydro-dynamical simulations of close encounters between pairs of BHs embedded in the gaseous disk and find that dissipation by gas gravitation is not always efficient for the formation of bound BBHs. Specifically, they find that it is usually an effective mechanism for BBH formation for gas densities $\rho_{g}$ in the range $-4.5 \lesssim \log\quadre{\rho_\mathrm{g}R_\mathrm{H}^{3}/\tonde{m_1+m_2} } \lesssim -2.5$, where 
$R_\mathrm{H}$ is the Hill radius of the BBH. In our simulations, we typically have $\log\quadre{\rho_\mathrm{g}R_\mathrm{H}^{3}/\tonde{m_1+m_2}} \simeq -6$ for our BBHs because the high gas density in the migration trap is compensated by small Hill radii due to the proximity to the SMBH, hence gas dissipation is expected to be inefficient. 

On the other hand, in their recent $N$-body simulations, \citet{Qian_2023} find that BBH formation from two BHs on similar orbits embedded in the disk is efficient for $\Omega\, \tau_0 \lesssim 10$, where $\Omega$ is the Keplerian angular velocity of the BHs and
\begin{equation}
    \tau_0 = \frac{c_\mathrm{s}^3}{4 \pi G \rho_{\rm g} m_1} 
    \label{eq:qian23}
\end{equation}
is the dynamical friction timescale \citep{Ostriker_1999} which, as shown in \citet{Qian_2023}, is linearly related to the timescale of BBH pair-up.
In our model typically $\Omega\, \tau_0 \sim 10^{-4}-10^{-3}$ due to the high gas density $\rho_{\rm g}$ in the migration trap, which justifies our assumption. 

Moreover, we expect a high number density $n_\mathrm{BH}$ of BHs in the migration trap due to efficient migration. This may further aid BBH pair-up 
because the timescales for two-body and three-body captures scale respectively as $n_\mathrm{BH}^{-1}$ \citep{quinlan-shapiro-1990} and $n_\mathrm{BH}^{-2}$ \citep{Fragione_t3bb}. In future work, we will further explore the process leading to BBH pair-up, accounting for these additional effects. 

In this work we neglect BBH formation in the bulk (i.e. outside of the migration trap), which is  expected to be efficient due to the high number of Type I migrators embedded in the disk \citep[e.g.][]{Tagawa_2020_b, Tagawa_2020}. Interestingly, 
gas-assisted bulk-assembled BBHs migrate toward the migration trap while hardening \citep[Fig. 8]{Tagawa_2020}. Hence, some binaries may merge in the migration trap even if they were assembled in the bulk.

\subsection{Three-body encounters} 
\label{sec:threebody}
\begin{figure}
    \centering
    \includegraphics[width=\linewidth]{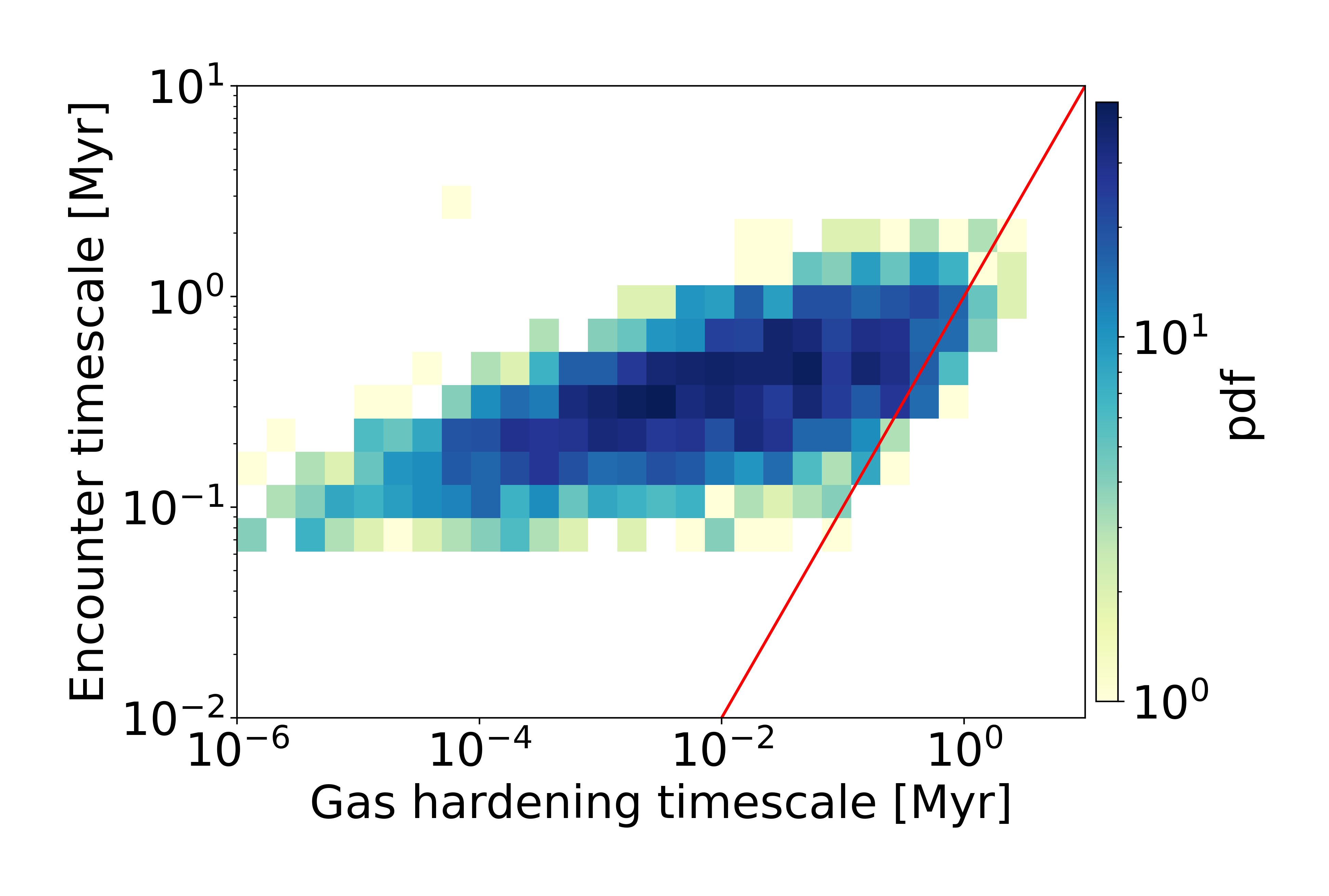}
    \caption{Delay timescale $t_\mathrm{del}$ in the GH model compared against the three-body encounter timescale $t_\mathrm{enc}$, estimated as in \citet{Leigh_McKernan_2018}. The red line is $t_\mathrm{del}=t_\mathrm{enc}$.}
    \label{fig:threebody_timescale}
\end{figure}
We neglect three-body interactions and their effects on BH scattering and BBH hardening. We make this 
approximation because of the dearth of semi-analytical models for these interactions: in the literature, there are some works (e.g., \citealt{Miller_t12, Fragione_t3bb}) that assume an isotropic distribution of velocities and are appropriate for spherical star clusters. In a Keplerian disk geometry, the distribution of velocities is much different and these models are not appropriate \citep{McKernan_2022}. 
Nevertheless, a hard binary hardens by binary-single encounters \citep{Heggie_hard_bin} and this has a two-fold effect: the inspiral of BBHs is accelerated by three-body effects, and the third intruding body receives a recoil kick which may prevent it from reaching the migration trap.
Also, binary-single interactions are expected to increase the eccentricity of all BBHs to $e\tonde{\SI{10}{\hertz}} \geq 10^{-4}$ and potentially flip the orientation of the BBH orbital angular momentum, leaving key signatures in the effective spin distribution \citep{Samsing_2022}. 
Additionally, \citet{Rowan_2023} show that the initial eccentricity of BBHs formed in AGN disks is high, which is not well modeled by our assumption of a thermal distribution \citep{Jeans_1919_binaries}.

\citet{Tagawa_2020} simulated the evolution of the compact object population in AGN disks using a one-dimensional N-body simulation combined with a semi-analytical model for the formation, disruption, and evolution of binaries. 
They include the effects of three-body interactions and employ a \citet{TQM} disk model rather than the \citetalias{SG} model that we use.
Nevertheless, comparing our Fig.~\ref{fig:mass-ratio} with \citet[Fig. 12a]{Tagawa_2020}, we can point out that the output from our fiducial model is compatible with the output from theirs. This suggests that three-body effects have a relatively minor impact on the population of merging BBHs. 

\citet{Leigh_McKernan_2018} find that the approximate encounter timescale between a BH in the migration trap and an intruder is a fraction of the Type I migration timescale (eq.~\ref{eq:t_migr}) of an object of mass $m_*$ starting its migration from the outer skirts of the disk, namely
\begin{equation}
    t_\mathrm{enc} \simeq \frac{t_\mathrm{migr,\,I}\tonde{R_\mathrm{max}}}{N_*} = \frac{1}{N_*} \quadre{ \frac{\MSMBH^2\,  h^2\tonde{R_\mathrm{max}}}{m_*\,{} \Sigma_\mathrm{gas}\tonde{R_\mathrm{max}} \,{}R_\mathrm{max}^2\,{} \Omega_*}} ,
\end{equation}
where $N_*$ is the number of Type I migrators in the disk. 

Assuming $m_*=1\Msun$ and $N_*=100$, we find that the encounter timescale is typically larger than the delay timescale in the GH scenario (Fig.~\ref{fig:threebody_timescale}), justifying our disregard of three-body effects in this scenario. Nevertheless, they should be included in the no-GH model.

This implies that Heggie's law (employed in eq.~\ref{eq:Heggie's law}) may not be valid in the GH case, as even soft binaries may be hardened considerably on a timescale shorter than the one on three-body interactions. We find that forming soft BBHs in our model is an extremely rare event (less than $1$ in $10000$ BBHs), so including them in our computation would not have affected the results significantly. 

Furthermore, efficient damping, migration, and BBH pair-up in AGN disks are expected to give rise to a high concentration of BBHs in migration traps. Hence, we should take into account the effect of binary-binary interactions. 
Such encounters are expected to efficiently ionize one of the binaries involved and lead to the formation of a stable triple \citep{pina-gieles2023}.

\subsection{Position of migration traps}
\label{sec:discussion_migration}
\begin{figure*}
    \centering
    \includegraphics[width=\textwidth]{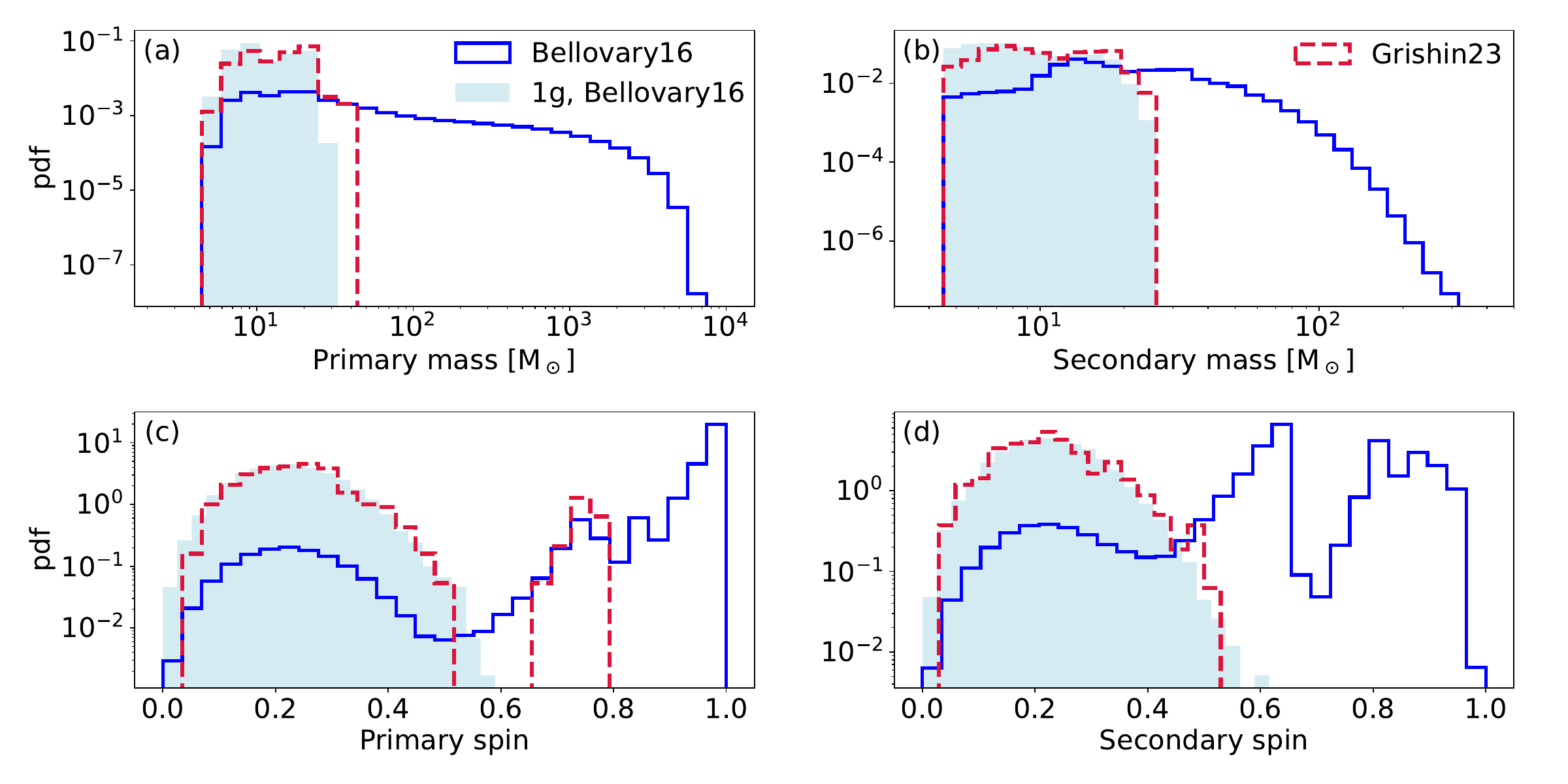}
    \caption{Main properties of dynamical BBH mergers with two different models for the migration timescale and the location of migration traps. Blue solid line: fiducial GH model. Red dashed line: we use the model by \citet{grishin2023} for both the migration timescale and the location of the migration trap. Light-blue filled histograms refer to $1g$ mergers in the fiducial model. \textbf{(a,b)} Primary and secondary BH masses. 
    \textbf{(c,d)} Primary and secondary BH spin magnitudes. 
    }
    \label{fig:Grishin23}
\end{figure*}
The position of migration traps is highly uncertain: it strongly depends on the disk model and the migration prescription used. For instance, \citet{Pan_2021} compute Type I migration torques in the locally-isothermal approximation as well as torques due to winds and to the gravitational interaction with the SMBH for three different disk models. In their approximation, no disk model develops migration traps. Moreover, they find that in the inner part of the disk ($R\lesssim 10^2 R_\mathrm{g}$) Type I migration torques are always overpowered by the other effects, which prompt embedded BHs to eventually merge with the SMBH. This justifies our assumption of neglecting the inner migration trap for the \citetalias{SG} model (Section~\ref{sec:setup_AGN_methods}), but it implies that we somewhat overestimate the number of Type I migrators.

In their recent work, \citet{grishin2023} account for both Type I migration, caused by the gravitational perturbation of embedded BHs in the disk, and thermal migration, caused by the thermal response of the disk to the small and overdense accretion disks surrounding embedded BHs. They find that the resulting migration trap position 
is at much larger distances than previously identified by considering Type I migration only.

In our fiducial model, we followed the prescription by \citet{Bellovary_2016} and identified the position of the migration trap as $R_\mathrm{trap} = 1324\, R_\mathrm{g}$,
whereas \citet{grishin2023} find that the updated position of the migration trap is typically  larger $R_\mathrm{trap} = 10^3-10^6\, R_\mathrm{g}$, and has a steep dependence on the SMBH mass as
\begin{equation}
    \log{\tonde{R_\mathrm{trap}/R_\mathrm{g}}} \simeq -\log{\tonde{\MSMBH/\Msun}} + 11,
\end{equation}
for $\MSMBH\lesssim 10^8 \Msun$. Hence, for a typical SMBH of mass $10^6 \Msun$, the migration trap is at $\sim 10^5\, R_\mathrm{g}$.

Migration is also affected by the additional thermal torque, which reduces the migration timescale by a factor of
\begin{equation}
    t_\mathrm{migr}^\mathrm{tot} \simeq \frac{h}{2}\, t_\mathrm{migr,\,I}, 
\end{equation}
where $t_\mathrm{migr,\,I}$ is defined as in eq. \ref{eq:t_migr}. 

Figure~\ref{fig:Grishin23} shows a comparison between the results of our fiducial model and 
the results we obtain assuming the position of migration traps and the migration timescales predicted by \citet{grishin2023}. 
We find that the hierarchical merger  process is 
suppressed when including a treatment for thermal torques, as only a handful of seed BHs reach the second generation in our simulations. This happens because the gaseous disk is significantly thicker and more dilute in the outer area of the disk where the new migration trap is located (see Section~\ref{sec:setup_AGN_methods}). Therefore, as the migration timescale is proportional to $h^3$ and $\Sigma_\mathrm{gas}^{-1}$, the migration process is much slower in this scenario. Moreover, even when BHs successfully reach the migration trap, the delay timescale 
ends up being too long because of the low gas surface density.

\subsection{General caveats}

In our model, we disregard dynamical interactions with the SMBH and gravitational perturbations caused by intermediate-mass BHs. For example, \citet{Deme_2020} show that the presence of intermediate-mass BHs in the disk may enhance the ionization of BBHs and consequently decrease the merger rate.

Moreover, in Section~\ref{sec:damping} we assumed that eccentricity and inclination of a BH  orbit are damped on similar timescales due to gas torques. \citet[Fig. 4]{wang2023} show that this is not always accurate: if the initial orbit is highly eccentric ($e\gtrsim0.9$), the eccentricity-damping timescale can be up to five times longer than the inclination-damping one. A BH on a gas-embedded eccentric orbit ($i\sim0$, $e\gtrsim h$) would be subject to spin-down \citep{McKernan_2023}, which would leave 
an imprint on the expected spin distribution.

Finally, we ignore the evolution of AGN disks in time. This can happen slowly over the AGN lifetime as BHs are embedded in the disk \citep[][]{Tagawa_2022} and gas is accreted by the SMBH. 
The efficiency of all dynamical processes
strongly depends on disk density and aspect ratio. If these quantities evolve over time, the resulting BBH population will be affected as well.

\subsection{GW190521 and other transient events}
It has previously been suggested that the transient events GW170729 and GW190521 could have originated in AGN disks \citep{Yang_2019, Tagawa_2021, Samsing_2022, Graham_gw190521, Morton_in_prep}. In particular, \citet{Yang_2019} found that it is 5 times more likely for GW170729 to arise from hierarchical mergers in AGNs than assuming that all events in the GWTC-1 catalog \citep{GWTC1_Abbott_2019} arise from the same channel, whereas \citet{Morton_in_prep} show that the association between GW190521 and the AGN flare ZTF19abanrhr \citep{Graham_gw190521} is highly preferred over the lack of association, suggesting that the GW transient event was generated in an AGN. 

We compare the posterior contours for the primary mass and effective spins of such events with the output of our AGN simulations, as shown in Fig.~\ref{fig:hist-LVK-comparison}. We also include the posterior contour of the transient event GW190403$\_$051519, although it has a low SNR of $7.6^{+0.6}_{-1.1}$ and low probability of astrophysical origin $p_\mathrm{astro}=0.61$ \citep{GWTC2.1}. 

We find that the contours of GW170729 have some overlap with both the GH and the no-GH AGN models, and thus might be compatible with an origin in an AGN environment. 
In contrast, GW190521 has no significant overlap with 
our models in the $\chi_\mathrm{eff}-m_1$ space, and negligible overlap with our no-GH model in the $\chi_\mathrm{eff}-\chi_\mathrm{p}$ space. Indeed, the posterior probability distribution of GW190521 shows moderate support for high precession spin and low effective spin, suggesting that its BH spin vectors are large but misaligned. In contrast, in our GH model, we predict that they should be aligned for a BBH of primary mass $\sim 100 \Msun$. This does not rule out the hypothesis of an AGN origin for GW190521, as we might speculate that the BBH may have been perturbed by three-body encounters,  which we do not account for in our GH model (Appendix~\ref{sec:spin_tilt}). This would alter the orientation of the orbital angular momentum and decrease the effective spin, as well as increase the BBH eccentricity \citep{Samsing_2022}. 

Finally, 
the posterior contours of GW190403$\_$051519 
significantly overlap with our GH AGN model. However, this event candidate has a high false alarm rate and may not have astrophysical origin \citep{GWTC2.1}.


\section{Summary} 
\label{sec:summary}

We explored the formation of binary black hole (BBH) mergers in active galactic nuclei (AGNs) employing a new semi-analytical model. We summarize our key findings as follows.
\begin{itemize}
    \item The presence of gas hardening (GH) significantly increases the efficiency of hierarchical mergers in AGN disks, allowing seed black holes (BHs) to go through up to a thousand merger episodes. This leads to the formation of BBHs with high mass (up to a few thousand solar masses), and low mass ratio $q\simeq 10^{-2}$.
    \item In contrast, if GH is not efficient, the hierarchical merger chain is truncated after a few generations, leaving a BBH population with no components more massive than $\sim 10^2 \Msun$.
    \item The distribution of spin tilt angles is influenced by GH, causing preferential alignment of spins in the GH scenario and isotropic distribution in the no-GH scenario. Hence, GH leads to distinct features in the effective spin distributions: a main peak on $\chi_\mathrm{eff}\simeq1$ corresponding to maximum alignment, and smaller peaks at $\chi_\mathrm{eff}\simeq0.2$ and $\chi_\mathrm{eff}\simeq 0.5$.
    \item In the GH scenario, we find an anti-correlation between $q$ and $\chi_{\rm eff}$ that might be extending even to lower values of $q$ and higher values of $\chi{}_{\rm eff}$ than currently observed by LVK.
    \item Comparison with other formation channels shows that AGN-driven mergers in the GH scenario result in higher primary BH mass and higher effective spin. 
\end{itemize}

In summary, we find that efficient gas hardening in AGN disks enhances the formation of BBH mergers with high primary mass and low mass ratios, and leads to a strong preference for $\chi_{\rm eff}\approx{1}$. Given the large BH mass, next-generation ground-based detectors like the Einstein Telescope are an ideal test bed for such unique features.

\section*{Data Availability}
\fastcluster{} is an open-source code available at \href{https://gitlab.com/micmap/fastcluster_open}{this link}. The latest public version of {\sc sevn} can be downloaded from \href{https://gitlab.com/sevncodes/sevn}{this repository}. The data underlying this article will be 
shared on reasonable request to the corresponding authors. 

\begin{acknowledgements}
We thank the anonymous referee for their constructive and insightful comments that improved this manuscript.
MM, MPV, CP, and ST acknowledge financial support from the European
Research Council for the ERC Consolidator grant DEMOBLACK,
under contract no. 770017. MPV and MM acknowledge financial support from the German Excellence Strategy via the Heidelberg Cluster of Excellence (EXC 2181 - 390900948) STRUCTURES. MD acknowledges financial support from the Cariparo Foundation under grant 55440. We thank Tamara Bogdanovi{\'c}, Barry McKernan,  Dominika Wylezalek, Jenny Greene, Carolin ``Lina'' Kimmig, Ralf Klessen, Gast{\'o}n Javier Escobar, Stefano Rinaldi, Giorgio Mentasti, Jacopo Tissino,  Imre Bartos, Bence Kocsis, K.E. Saavik Ford, and Hiromichi Tagawa for their useful input. We thank Dylan Nelson for granting us access to the IllustrisTNG JupyterLab workspace. 
\end{acknowledgements}

%
\bibliographystyle{aa} 
\bibliography{bibliography.bib} 
%

\begin{appendix}
\section{Catalogs for AGN disks, NSCs, GCs, YSCs, and isolated binaries}
\label{sec:oldfastcluster}
We generated catalogs of BH masses for AGNs, NSCs, GCs, YSCs, and isolated binaries with the {\sc sevn} binary population synthesis code \citep{sevn_2023}. For the isolated binary systems, we have integrated the evolution of $7.5\times{}10^7$ binary systems divided in 15 metallicity bins ranging from $Z=10^{-4}$ to $3\times{}10^{-2}$. The initial properties of such binary systems and the treatment of binary evolution processes (stable mass transfer, common envelope, tides, natal kicks, gravitational-wave decay, etc) are the same as in the fiducial model by \cite{sevn_2023}. Also, we consider the rapid model for core-collapse supernovae \citep{Fryer_2012} and include a treatment for pair instability as described by \cite{sevn_2020}.

The masses of BHs in AGNs, NSCs, GCs, and YSCs are obtained from a single star population run with {\sc sevn}. In particular, for NSCs, GCs, and YSCs we simulated the same set of 15 different metallicities ranging from $Z=10^{-4}$ to $3\times{}10^{-2}$ as in the isolated binary case but we turn off binary evolution. These catalogs are the same as we adopt for AGN BHs, but for the latter we only used the solar metallicity $Z=0.02$, because the metallicity at the center of a massive galaxy tends to be solar or super-solar. In future work, we will also model the metallicity evolution in AGN disks. 

The spin magnitudes of BHs in both isolated binaries and star clusters are drawn from a Maxwellian distribution with $\sigma{}_\chi{}=0.1$, the same as for BHs in AGN disks. In the dynamical channel, spin orientations are assumed to be isotropic, because dynamical encounters randomize them with respect to the orbital plane; whereas for isolated binary systems we derive the final spin orientation accounting for the effect of natal kicks on the orbital plane \citep[e.g.,][]{Mapelli2021_review}.

BHs in NSCs, GCs, and YSCs pair up dynamically, via three-body and binary-single encounters, as described by \cite{fastcluster2021}. Here, we do not consider the contribution of primordial binaries to BBH mergers in NSCs, GCs, and YSCs \citep[see, e.g.,][for a treatment of primordial binaries]{fastcluster2021}. Subsequently, dynamically formed BBHs evolve via three-body hardening and gravitational-wave decay in their parent star clusters. When two BHs merge, we calculate the properties of the compact remnant (mass and spins) and its gravitational recoil. If the BH remnant remains in the star cluster, it can form a second-generation BBH and undergo further mergers (see \citealt{fastcluster2021} for more details). 
We estimate the 
evolution of BBH mergers in isolated binaries, NSCs, GCs, and YSCs as described by \cite{fastcluster2022} and summarized in Appendix~\ref{sec:cosmorate}.

\section{Spin tilt}
\label{sec:spin_tilt}

\subsection{Gas-hardening (GH) scenario}

Embedded objects can weakly perturb the surface-density profile of the AGN gaseous disk, resulting in gas torques that tend to align both the BH spin vectors $\vec\chi_{1,2}$ and the binaries’ orbital angular momentum vector $\vec L$ with the angular momentum $\vec J$ of the AGN disk \citep{Lubow_2015,Vajpeyi_2022}.

According to \citet{Bogdanovic_2007}, this process is particularly efficient if fully-embedded BHs can accrete more than $1\%$ of their mass due to gas accretion, so that $\Delta m_\mathrm{BH} \geq 0.01\, m_\mathrm{BH}$. 
Since 
we only consider BHs that are able to reach the migration trap and form a BBH, we 
estimate the mass variation of the binary as
\begin{equation}
    \Delta m_\mathrm{BBH} = \Delta m_1 + \Delta m_2 \simeq 
    \Sigma^\mathrm{trap}_\mathrm{gas} \pi \,{}a^2
\end{equation}
where $m_{1,2}$ are the primary and secondary BH masses, $a$ is the BBH semi-major axis, and the superscript "trap" indicates that the quantities are computed at the radial location $R_\mathrm{trap}$. Then, we check that 
\begin{equation}
    \Delta m_\mathrm{BBH} \geq 0.01 \tonde{m_1+m_2}.
\end{equation}
This condition is always verified in our model since the migration trap is the location with the highest surface density $\Sigma_\mathrm{gas}$ in the disk. Hence we model the spin tilt in our fiducial model as in the high-alignment scenario of \cite{Vajpeyi_2022}.

We sample $\cos{(\theta_1)}$ ($\theta_1$ being the angle between $\vec\chi_1$ and $\vec L$) from a truncated Gaussian centered in 1 with standard deviation $\sigma_1=0.1$, 
so that the primary spin is aligned with the orbital angular momentum. We sample $\cos{(\theta_2)}$ ($\theta_2$ being the angle between $\vec\chi_2$ and $\vec L$) from a truncated Gaussian centered in $\cos{\theta_1}$ with standard deviation $\sigma_2=0.1$, 
so that the spin of the secondary is aligned with that of the primary. 
This also implies $\vec\chi_2 \parallel \vec L$.
For the azimuthal direction, we draw $\cos{\phi_{1,2}}$ from a uniform distribution.

We point out that alignment (and not anti-alignment) is efficient in AGNs because, as shown by \cite{King_2005}, counter-alignment is only possible if $ J < 2\, \chi$, 
but in AGN disks $J$ is typically very large because the gas is in Keplerian motion close to a massive SMBH, so we can safely assume that the counter-alignment condition is never met. Moreover, retrograde binaries (i.e. with $\vec L\cdot \vec J <0$) are preferentially ionized or softened by tertiary encounters compared to prograde BBHs \citep{Wang_2021}.


\subsection{No gas-hardening (no-GH) scenario}

Gas torques are not the only physical phenomenon influencing BHs spin tilt: gas turbulence \citep{Chen_2023} and three-body encounters of BBHs with other objects \citep{Tagawa_2020_b} tend to randomize the alignment of $\vec\chi_{1,2}$ relative to $\vec L$. The competing effects of the gaseous disk and dynamical encounters on BBHs determine the distribution of BBH spin orientations. We neglect three-body encounters in our fiducial model because gas hardening significantly speeds up BBH  inspiral, so that the timescale $t_\mathrm{del}$ between BBH pair-up and merger (Section~\ref{sec:bbh_hardening}) is typically shorter than the typical timescale for three-body encounters $t_\mathrm{3bb}$ (which we estimate as in \citet[eq. 19]{Leigh_McKernan_2018}).
Instead, when we neglect gas hardening, the two timescales become comparable to each other. We assume three-body encounters to be the dominant effect in this scenario, similarly to what has been shown by \cite{Tagawa_2020_b} who also neglect gas hardening. In this model, we set the spin tilts $\theta_{1,2}$ from a Gaussian distribution (as we have said before), but with 
standard deviation $\sigma_1 = \sigma_2 = 10$, as in the isotropic scenario of \cite{Vajpeyi_2022}. Indeed, setting $\sigma_i=10$ is equivalent to sampling $\cos{\theta_i}$ from a uniform distribution in the interval $\quadre{0\,,1}$ and coincides with the isotropic case.
Once again, we draw $\cos{\phi_{1,2}}$ from uniform distributions.

\section{Secondary BH  mass}
\label{sec:secondary mass}


In our model, we allow for $Ng-Mg$ mergers, where $N$ and $M$ are the generation of the primary and secondary BH mass, respectively. 
Following \cite{Zevin-Holz}, we assume that the probability that a given generation $M$ is chosen for the secondary is proportional to the number of $1g$ BHs required to produce it: 
\begin{equation}
    p\tonde{M} \propto 2^{-(M-1)}
    \label{eq: prob gen M}
\end{equation}
for $M\leq N$, so that a $1g$ primary BH ($N=1$) will necessarily pair up with a $1g$ secondary ($M=1$). For $M=1$, the secondary BH mass $m_2$ is randomly drawn with probability distribution \citep{O_Leary_2016}
\begin{equation}
    p\tonde{m_2\,|\,m_1}\propto \tonde{m_1+m_2}^4
    \label{eq:o'leary}
\end{equation}
between $m_2^\mathrm{min}=5\Msun$ and $m_2^\mathrm{max}=m_1$.
 
For $M>1$, we determine the secondary mass as follows. First of all, we generate a $1g$ BH determining its mass and spin as in eq. \ref{eq:o'leary}. Then, we let it go through a certain number of $Ng-1g$ mergers\footnote{We choose to consider only $Ng-1g$ mergers rather than $Ng-Mg$ for simplicity.} until it creates an $Mg$ remnant. 

At each step, the primary will be the remnant of a previous merger event 
and 
its mass $m_1$ and spin will be computed according to \citet{Jimenez}. The secondary BH mass will be sampled from 
\begin{equation}
    p(m_2\,|\,m_1)\propto (m_1+m_2^\mathrm{max})^4
    \label{eq:modified o'leary}
\end{equation}
between $m_2^\mathrm{min}=5 \Msun$ and $m_2^\mathrm{max}$.

Here, $m_2^\mathrm{max}$ is determined based on the value of $m_1$. In particular, if $m_1$ is smaller than the maximum mass $m_1^\mathrm{max}$ of a $1g$ BH in the input sample coming from the population synthesis code \textsc{sevn}, the secondary mass cannot exceed the mass of the primary: $m_2^\mathrm{max}=m_1$. Otherwise, we set $m_2^\mathrm{max}=m_1^\mathrm{max}$. 
This is a modification of eq. \ref{eq:o'leary}.

With this choice, if the primary has mass compatible with a $1g$ BH ($m_1 \leq m_1^\mathrm{max}$) we sample the secondary as seen previously in eq. \ref{eq:o'leary} (model from \citealt{O_Leary_2016}).

Otherwise, if the primary is a higher-generation BH ($m_1 > m_1^\mathrm{max}$), we keep the same analytical description as in eq. \ref{eq:o'leary} but we force the secondary to have mass compatible with a $1g$ BH ($m_2 \leq m_1^\mathrm{max}$).
This calculation is quite fast because we only compute the remnant mass and spin at each step, neglecting the merger times.

\section{COSMO\texorpdfstring{$\mathcal{R}$}{R}ATE}
\label{sec:cosmorate}
We interface the catalogs produced in our computation with the code \textsc{cosmo}$\mathcal{R}$\textsc{ate} \citep{Santoliquido_2020,Santoliquido_2021}, which calculates the BBH merger rate evolution $\mathcal{R}\tonde{z}$ by using catalogs of BBH mergers simulated with \fastcluster{} and by coupling
them with the cosmic star formation rate $\psi\tonde{z}$ and metallicity evolution as
\begin{equation}
    \mathcal{R}\tonde{z} = \int_{z_\mathrm{max}}^z \psi\tonde{z'}\frac{\mathrm{d}t\tonde{z'}}{\mathrm{d}z'} \quadre{\int_{Z_\mathrm{min}}^{Z_\mathrm{max}} \eta\tonde{Z} \mathcal{F} \tonde{z',z,Z} \mathrm{d}Z} \mathrm{d}z' ,
\end{equation}
where $t\tonde{z'}$ is the lookback-time at redshift $z'$ and $\mathrm{d}t\tonde{z'}/\mathrm{d}z'=\quadre{\tonde{1+z'} H\tonde{z'}}^{-1}$, with $H\tonde{z'}=H_0\quadre{\tonde{1+z'}^3 \Omega_\mathrm{M}+\Omega_\Lambda}^{1/2}$.

For the AGN channel, we assume solar metallicity $Z=0.02$. We compute the fraction $\mathcal{F}\tonde{z',z,Z}$ of BBH that form at redshift $z'$ and merge at redshift $z$ by interpolating the density $n_{\rm AGN}\tonde{z'}$ of active SMBHs with $\dot M \geq 0.2 \dot M_{\rm Edd}$, shown in Fig.~\ref{fig:activenum/comovingvol}, and keeping into account the merger timescale $t_\mathrm{merg}$. We compute the merger efficiency $\eta\tonde{Z}$ as the ratio between the total number of BBH mergers with merger timescale lower than the Hubble timescale ($t_\mathrm{merg}\leq \SI{13.4}{\giga\year}$) and the total simulated stellar mass (eq.~\ref{eq:Plummer integrated}). We repeat the computation for other \fastcluster{} channels as illustrated in  \citet{fastcluster2022}. We assume metallicity spread $\sigma_Z = 0.3$.

The \textsc{cosmo}$\mathcal{R}$\textsc{ate} algorithm can 
produce catalogs of BBH mergers at different redshifts. In Section~\ref{sec:multichannel} we compare the outputs of five channels at low redshift $z\leq0.1$. The full five-channel catalog up to redshift $z=15$ is used for the computation of mixing fractions illustrated in Appendix~\ref{sec:mix_frac}.

\section{Mixing fractions}
\label{sec:mix_frac}
We compare our models against the 56 high-purity GW events 
analyzed by \citet{GWTC3_second} using a hierarchical Bayesian approach as in \citet{fastcluster2022}. We shortly describe the process here for the reader's comfort.

Given a number $N_\mathrm{obs}$ of GW observations $\mathcal{H}=\graffe{h_k}_{k=1}^{N_\mathrm{obs}}$ described by an ensemble of parameters $\theta$, the posterior distribution of the hyper-parameters $\lambda$ associated with the models is
\begin{equation}
    p\tonde{\lambda,N_\lambda\,|\,\mathcal{H}} = e^{-\mu_\lambda} \pi(\lambda,N_\lambda) \prod_{k=1}^{N_\mathrm{obs}} N_\lambda \int_\theta \mathcal{L}^k (h^k\,|\,\theta) p(\theta\,|\,\lambda) \,d\theta ,
    \label{eq:posterior-hyper-param}
\end{equation}
where $\theta=\graffe{M_\mathrm{chirp}, m_1+m_2,\chi_\mathrm{eff},\chi_\mathrm{p},z}$ are the GW parameters, $N_\lambda$ is the number of events predicted by the astrophysical model, $\mu_\lambda$ is the predicted number of detections associated with the model and the GW detector, $\pi(\lambda, N_\lambda)$ is the prior distribution on $\lambda$ and 
$N_\lambda$, and $\mathcal{L}^k (h^k\,|\,\theta)$ is the likelihood of the $k$-th detection. The predicted number of detections is given by
\begin{equation}
    \mu_\lambda = N_\lambda \int_\theta p(\theta\,|\,\lambda) p_\mathrm{det}(\theta) \,d\theta ,
\end{equation}
where $p_\mathrm{det}$ is the probability of detecting a source with parameters $\theta$ and can be inferred by computing the optimal signal-to-noise ratio and comparing it to a detection threshold, as described in \citet{Bouffanais_2021b}. Also, we marginalize over $N_\lambda$ using a prior $\pi(N_\lambda) \sim 1/N_\lambda$ \citep{Fishbach_2018}. This implies that our mixing fractions do not depend on the merger rate of each channel.

We compute the mixing fractions $\graffe{f_\mathrm{iso}, f_\mathrm{AGN}, f_\mathrm{GC}, f_\mathrm{YSC}, f_\mathrm{NSC}}$ which weight the contributions of our five channels 
to the overall distribution ($\rm i=isolated, AGN, NSC, GC, YSC$):
\begin{equation}
p\tonde{\theta\,|\,\lambda} = \sum_{\rm i} f_\mathrm{i}\, p\tonde{\theta\,|\,\mathrm{i},\lambda} ,
\end{equation}
The mixing fractions are defined so that 
\begin{equation}
    f_\mathrm{iso}+ f_\mathrm{AGN}+ f_\mathrm{GC}+ f_\mathrm{YSC}+ f_\mathrm{NSC}=1 .
\end{equation} 
Based on this definition, the mixing fraction for each channel is approximately the fraction of merger events associated with that specific channel. 
This definition of the mixing fraction assumes that all GWTC-3 events
originate from the five channels we considered here, so we are neglecting the possibilities of BBH mergers from primordial BHs, triples, and multiples as well as any other possible evolution channel.

\end{appendix}

\end{document}